\documentclass[12pt]{article}
\usepackage{latexsym}
\usepackage{amsmath,amsbsy,amssymb}
\usepackage{verbatim}
\usepackage{bm}
\usepackage{epsfig}

\newcommand{\dd}{\mbox{\rm d}}

\newcommand{\gam}{\gamma}

\newcommand{\dg}{\dagger}

\newcommand{\tl}{\tilde}
\newcommand{\ul}{\underline}

\newcommand{\dotx}{{\dot x}}
\newcommand{\ddotx}{{\ddot x}}
\newcommand{\om}{{\omega}}

\newcommand{\nnn}{\noindent}

\newcommand{\p}{\partial}

\newcommand{\be}{\begin{equation}}
\newcommand{\bear}{\begin{eqnarray}}
\newcommand{\ear}{\end{eqnarray}}
\newcommand{\ee}{\end{equation}}

\newcommand{\lbl}{\label}
\newcommand{\bi}{\bibitem}
\newcommand{\ci}{\cite}

\newcommand{\vs}{\vspace}
\newcommand{\hs}{\hspace}

\newcommand{\bp}{{\bm p}}

\newcommand{\ba}{{\bar a}}

\begin{document}

\begin{center}

\

\vs{.8cm}

\baselineskip .7cm

{\bf \Large Pais-Uhlenbeck Oscillator and Negative Energies} 

\vs{4mm}

\baselineskip .5cm
Matej Pav\v si\v c

Jo\v zef Stefan Institute, Jamova 39,
1000 Ljubljana, Slovenia

e-mail: matej.pavsic@ijs.si

\vs{3mm}

{\bf Abstract}
\end{center}

\baselineskip .45cm

We review the occurrence of negative energies in Pais-Uhlenbeck oscillator. We point out
that in the absence of interactions negative energies are not problematic, neither in the
classical nor in the quantized theory. However, in the presence of interactions
that couple positive and negative energy degrees of freedom the system is unstable,
unless the potential is bounded from bellow and above.
We review some approaches in the literature that attempt to avoid the problem of negative
energies in the Pais-Uhlenbeck oscillator.

\vs{2mm}


\baselineskip .55cm

\section{Introduction}

Higher derivative theories could be one of possible roads to quantum gravity.
But the wrong sign in the Ostrogradsky Hamiltonian\,\ci{Ostrogradski} of such
theories has prevented
them to be considered as viable physical theories.  A toy model for higher-derivative theories
is the Pais-Uhlenbeck (PU) oscillator\ci{PaisUhlenbeck}.  It has been a common believe that, because of
the presence of negative energy states, the PU oscillator is unstable. In the absence
of interaction, negative energies are not
problematic\,\ci{PavsicPseudoHarm,Woodard}. But a completely isolated
physical system is unrealistic, therefore even a small interaction causes the transitions
between positive and negative energies.  If the interaction potential is unbounded,
like in harmonic oscillator, then it causes a runaway behaviour of the system. 
But an unbounded potential, either from below or form above, is physically unrealistic.
We know that the potential of a harmonic oscillator is an idealization; there is no such a
potential in the nature. Therefore, in the case of the PU oscillator, or any other system
that has negative energy states, a realistic potential is bounded from above and from below.
In such a case, the system is stable in the sense that the runaway behavior with
increasing positive energies (and decreasing negative energies) cannot superced
the height of the potential\,\ci{PavsicPUstable}.

In this paper I review how various authors have attacked the problem of negative
energies in PU oscillators.  We focus to those works in which the original Pais-Uhlenbeck
theory is assumed, and is not generalized to its non-Hermitian, but PT
symmetric version\,\ci{Mannheim}--\ci{Bender3}.
In most approaches with the original PU Lagrangian\,\ci{Mostafazadeh}--\ci{Masterov:2016jft}
(see also \ci{Dector}) the Hamiltonian
of the {\it non interacting} PU oscillator is transformed into a form in which only positive energies
occur. Those authors say nothing about how such procedures would work in the case
of an interacting or self-interacting PU oscillator. In Ref.\,\ci{PavsicPUstable} it was shown
that though a free PU oscillator Lagrangian can be transformed into a form which leads
to positive energies only, this cannot be done for a self-interacting PU oscillator with a quartic
interaction term. In the presence of an interaction term, a system with indefinite
Hamiltonian cannot behave as a system with positive definite Hamiltonian.
The approaches of Refs.\,\ci{Mostafazadeh}--\ci{Nucci} and also of\,\ci{Ghosh} 
thus reconfirm the known
fact\,\ci{PavsicPseudoHarm,Woodard,PavsicPUstable} that a free
system with indefinite Hamiltonian behaves just like a system with positive definite Hamiltonian.
Therefore, in an attempt to solve the negative energy problem, one must consider the
PU oscillator with an interaction term right from the beginning. 
The investigations by Pagani et al.\,\ci{Pagani} showed that such a system can be
stable for certain interaction potentials and choices of parameters.
Smilga\,\ci{Smilga1}--\ci{SmilgaStable} has
found that the classical interacting PU oscillator is stable  for certain range of initial data.
Such a system would still not be stable after quantization because of the tunnel effect. 
But in Ref.\,\ci{Robert} an unconditionally stable interacting system was found.
Further, in Ref.\,\ci{PavsicPUstable} an example the PU oscillator with an interaction term
bounded from above and below, given by the forth power of sine, was considered and shown
that such a system is stable for all initial conditions. The Pais-Uhlenbeck oscillator as
a fourth order system in derivatives with respect to time can be written as a system of
two oscillators with equal masses. If the masses of the the oscillators in such a
system are different, then the system is stable even in the presence of a coupling
term\,\ci{PavsicFirenze,PavsicPUstable}. 

Usually, when speaking about higher derivative systems, it is stated that upon
quantization such theories contain ghosts, and therefore violate unitarity. However,
as shown in Refs.\,\ci{Kim,Cangemi,Benedict,PavsicPseudoHarm,Woodard}, whether in a
quantized theory we have negative probabilities (ghosts) and positive energies,
or vice versa, positive probabilities and negative energies, depends on choice of vacuum.
Woodard\,\ci{Woodard} was very explicit in saying that a correct quantization should
always have positive probabilities, while the energies can be negative, otherwise
the correspondence principle is not satisfied.

Another important issue had been raised by Nesterenko\,\ci{Nesterenko} who considered
the effect of damping on the PU oscillator. He found that in the presence of damping the PU oscillator
is unstable. This was further explored in Ref.\,\ci{PavsicPUdamp} where it was found that
the instability does not occur if besides a damping term with the first order derivative
we have also a term with the third order derivative. In the presence of both terms
the system becomes stable.

Finally, it is pointed out how the issue of instantaneous vacuum decay in higher derivative field theories
can be clarified.

\section{Pais-Uhlenbeck oscillator as a toy model for higher derivative theories}

Whereas the theory based on the Einstein-Hilbert action is not renormalizable, the $R+R^2$ gravity
is renormalizable. In spite of this nice property, such a theory, because of the presence of ghosts,
has been generally considered as problematic and hence dismissed by majority of researchers.
The r\^ ole  of ghosts can be studied on a simpler example, namely, the fourth-order scalar
field theory with the action \ci{Hawking-Hertog}
\be
  I = \int \dd^4 x \left [\frac{1}{2} \phi(\Box +m_1^2)(\Box +m_2^2)\phi - \lambda \phi^4     \right ].
\lbl{2.1}
\ee
Defining new variables\,\ci{Hawking-Hertog}
\be
  \psi_1 = \frac{(\Box +m_2^2)\phi}{[2 (m_2^2-m_1^2)]^{1/2}}~,~~~~~ 
  \psi_2 = \frac{(\Box +m_1^2)\phi}{[2 (m_2^2-m_1^2)]^{1/2}} ,
\lbl{2.2}
\ee
where $m_2 > m_1$, the action (\ref{2.1}) becomes
\be
  I = \int \dd^4 x  \left [ \frac{1}{2} \psi_1 (\Box +m_1^2) \psi_1 -
   \frac{1}{2} \psi_2 (\Box +m_2^2) \psi_2 - \frac{4 \lambda}{(m_2^2-m_1^2)^2} (\psi_1 - \psi_2)^4 \right ].
\lbl{2.3}
\ee
This gives the following coupled equations of motion
\bear
 &&(\Box +m_1^2) \psi_1 - \frac{16 \lambda}{(m_2^2-m_1^2)^2} (\psi_1 - \psi_2)^3 = 0 \lbl{2.4}\\
&&(\Box +m_2^2) \psi_2 - \frac{16 \lambda}{(m_2^2-m_1^2)^2} (\psi_1 - \psi_2)^3 = 0 \lbl{2.5}
\ear
In the absence of interactions, there would be no problem. We would just have two independent,
uncoupled equations. The problem arises if $\lambda \neq 0$. Then  the energy flows between
the two fields and supposingly causes a runaway behaviour of the system.

In order to facilitate the study, instead of the field action (\ref{2.1}), one suppresses the spatial
dependence of the fields and considers the action
\bear
  I &&= \int \dd t \, \left [ \frac{1}{2} \phi \left ( \frac{\dd^2}{\dd t^2}+m_1^2 \right )
  \left ( \frac{\dd^2}{\dd t^2}+m_2^2 \right )\phi - \lambda \phi \right ] \nonumber \\
  &&= \int \dd t \left [ \frac{1}{2} {\ddot \phi}^2 - \frac{1}{2} (m_1^2 + m_2^2){\dot \phi}^2
  +\frac{1}{2} m_1^2 m_2^2 \phi^2 - \lambda \phi^4 \right ] +
  {\rm boundary ~term} .
\lbl{2.6}
\ear
This is the action for a higher derivative harmonic oscillator, in  the literature known as
Pais-Uhlenbeck oscillator\,\ci{PaisUhlenbeck}. Notation is usually adapted to such a simplified
system and the following Lagragian is considered:
\be
  L=\frac{1}{2} \left [ {\ddot x}^2 - (\om_1^2 + \om_2^2) \dotx^2 + \om_1^2 \om_2^2 x \right ]
  -\frac{\Lambda x^4}{4}
\lbl{2.7}
\ee

According to the Ostrogradski second order fromalism we have
\be
  p = \frac{\p L}{\p \dotx}=(\om_1^2 + \om_2^2) \dotx~,~~~~~P= \frac{\p L}{\p \ddotx} = \ddotx
\lbl{2.8}
\ee
\bear
  &&H= p \dotx + P \ddotx - L\nonumber\\
  &&\hs{5mm}= p_x q + \frac{1}{2}\left [p_q^2 + ( \om_1^2 + \om_2^2)q^2 - \om_1^2 \om_2^2 x \right ]
  +\frac{\Lambda x^4}{4} .
\lbl{2.9}
\ear
Because the momentum $p_x$ occurs linearly in (\ref{2.9}) the Hamiltonian is not positive
definite. It can have positive or negative values. This is a manifestation of the fact that the
PU oscillator possesses the so called {\it Ostrogradski instability}. Much effort has been
devoted in attempts to circumvent  this problem.

\section{Non interacting Pais-Uhlenbeck oscillator}

If the coupling constant $\lambda$ is zero, then the equation of motion is
\be
   x^{(4)} + (\omega_1^2 + \omega_2^2) {\ddot x} + \omega_1^2 \omega_2^2 x = 0.
\lbl{3.1},
\ee
which can be written as
\be
   \left ( \frac{\dd^2}{\dd t^2} + \omega_1^2 \right ) 
   \left (\frac{\dd^2}{\dd t^2} + \omega_2^2 \right ) x = 0,
\lbl{3.2}
\ee
In Refs.\,\ci{Bolonek,Mostafazadeh, Nucci} it has been shown that Eq.\,(\ref{3.1}) can be
derived from a positive definite Hamiltonian. An equivalent procedure was considered
by Stephen\,\ci{Stephen}, who distinguishes between Hamiltonian and the
``pseudo-mechanical energy". The procedure by Stephen and Mostafazadeh is
employed and generalized in Ref.\,\ci{PavsicPUstable} by using a different notation
as follows.

The fourth order equation (\ref{3.1}) can be written as a system of two second
order equations
\be
   {\ddot x} + \mu_1 x - \rho_1 y = 0,
\lbl{3.3}
\ee
\be
    {\ddot y} + \mu_2 y - \rho_2 x = 0,
\lbl{3.4}
\ee
provided that the real constants $\mu_1,~\mu_2,~\rho_1,~\rho_2$ satisfy the
relations
\be
     \mu_1 +\mu_2 = \omega_1^2 +\omega_2^2
\lbl{3.5}
\ee
\be
    \mu_1 \mu_2 - \rho_1 \rho_2 = \omega_1^2 \omega_2^2 .
\lbl{3.6}
\ee   
The solution is
\be
    \omega_{1,2}^2 = \mbox{$\frac{1}{2}$} (\mu_1 +\mu_2) \pm \mbox{$\frac{1}{2}$} 
    \sqrt{(\mu_1 +\mu_2)^2 - 4 (\mu_1 \mu_2 - \rho_1 \rho_2)} .
\lbl{3.7}
\ee

Eqs.\,({\ref{3.3}),(\ref{3.4}) can be derived from two different types of Lagrangians,
one associated a positive definite Hamiltonian, and the other one with an
indefinite Hamiltonian.

{\it Case I.} One possible Lagrangian is
\be
    L=\mbox{$\frac{1}{2}$}({\dot x}^2 + {\dot y}^2) - \mbox{$\frac{1}{2}$}
    (\mu_1 x^2 + \mu_2 y^2 - 2 \rho_1 x y ),
\lbl{3.8}
\ee
which gives the equations of motion ({\ref{3.3}),(\ref{3.4}) if $\rho_2 = \rho_1$.
Then Eq.\,(\ref{3.7}) reads
\be
    \omega_{1,2}^2 = \mbox{$\frac{1}{2}$}(\mu_1 + \mu_2) \pm \mbox{$\frac{1}{2}$} 
    \sqrt{(\mu_1 -\mu_2)^2 + 4  \rho_1^2} .
\lbl{3.9}
\ee
The latter relation admits real frequencies $\om_1$, $\om_2$ and thus oscillatory motion.

The second term in (\ref{3.8}) can be diagonalized if we perform a rotation in the
$(x,y)$-space,
\bear
    &&~x'=x\, {\rm cos}\, \alpha + y\, {\rm sin}\, \alpha \nonumber\\
    && y' = -x\, {\rm sin}\, \alpha + y\, {\rm cos}\, \alpha  ,
\lbl{3.10}
\ear
such that
\be
  \mu_1 x^2 + \mu_2 y^2 - 2 \rho_1 x y = a x'^2 + b y'^2 .
\lbl{3.11}
\ee
This gives the system of equations
\bear
     && a\, {\rm cos}^2 \,\alpha + b\, {\rm sin}^2 \, \alpha = \mu_1 \nonumber\\
     && a\, {\rm sin}^2 \,\alpha + b\, {\rm cos}^2 \, \alpha = \mu_2 \lbl{3.12}\\
     && (a-b)\, {\rm cos}\, \alpha \,{\rm sin}\, \alpha = \rho_1 \nonumber ,
\ear
whose solution is
\be
    a = \mbox{$\frac{1}{2}$}(\mu_1 + \mu_2) + \mbox{$\frac{1}{2}$} 
    \sqrt{(\mu_1 -\mu_2)^2 + 4  \rho_1^2} = \omega_1^2,
\lbl{3.13}
\ee
\be
    b = \mbox{$\frac{1}{2}$}(\mu_1 + \mu_2) - \mbox{$\frac{1}{2}$} 
    \sqrt{(\mu_1 -\mu_2)^2 + 4  \rho_1^2} = \omega_2^2,
\lbl{3.14}
\ee
\be
     {\rm cos}\, 2 \alpha= \frac{\mu_1-\mu_2}{\sqrt{(\mu_1 -\mu_2)^2 + 4  \rho_1^2}}.
\lbl{3.14a}
\ee

In the new coordinates the Lagrangian is thus
\be
    L=\mbox{$\frac{1}{2}$}({\dot x}'^2 + {\dot y}'^2) - \mbox{$\frac{1}{2}$}
    (\omega_1^2 {x'}^2 + \omega_2^2 {y'}^2),
\lbl{3.15}
\ee
the corresponding Hamiltonian being
\be
    H=\mbox{$\frac{1}{2}$}({\dot x}'^{\,2} + {\dot y}'^{\,2}) + \mbox{$\frac{1}{2}$}
    (\omega_1^2 {x'}^2 + \omega_2^2 {y'}^2).
\lbl{3.15a}
\ee
The energy os such a system is always positive. Let us mention that the system
described by the Lagrangian (\ref{3.15}) is equivalent to the PU oscillator
if $\om_1^2 \neq \om_2^2$. The case $\om_1^2 = \om_2^2$ is degenerate,
because then $\rho_1 = 0$, $\rho_2 =0$, and the system of equations
(\ref{3.3}),(\ref{3.4}) does not give the equation of motion of the PU oscillator.
Such a degenerated case has been considered in Ref.\,\ci{Ilhan}.

{\it Case II.} Another possible Lagrangian is
\be
    L=\mbox{$\frac{1}{2}$}({\dot x}^2 - {\dot y}^2) - \mbox{$\frac{1}{2}$}
    (\mu_1 x^2 - \mu_2 y^2 - 2 \rho_1 x y ),
\lbl{3.16}
\ee
which gives the equations of motion (\ref{3.3}),(\ref{3.4}) if $\rho_2 = - \rho_2$.
Then Eq.\,(\ref{3.7}) becomes
\be
    \omega_{1,2}^2 = \mbox{$\frac{1}{2}$}(\mu_1 + \mu_2) \mp \mbox{$\frac{1}{2}$} 
    \sqrt{(\mu_1 -\mu_2)^2 - 4  \rho_1^2} .
\lbl{3.17}
\ee
The frequencies $\om_1$, $\om_2$ are real if $(\mu_1 - \mu_2)^2 > 4 \rho_1^2$,
and $\mu_1 + \mu_2 > \sqrt{(\mu_1 -\mu_2)^2 - 4  \rho_1^2}$.

Now the second term in the Lagrangian (\ref{3.16}) can be diagonalized if we perform
the hyperbolic rotation in the $(x,y)$-space,
\bear
    &&x'=x\, {\rm cosh}\, \alpha + y\, {\rm sinh}\, \alpha \nonumber\\
    && y' =x\, {\rm sinh}\, \alpha + y\, {\rm cosh}\, \alpha , 
\lbl{3.18}
\ear
such that
\be
  \mu_1 x^2 - \mu_2 y^2 - 2 \rho_1 x y = a x'^2 - b y'^2 ,
\lbl{3.11}
\ee
which gives the system of equations
\bear
     && \omega_1^2\, {\rm cosh}^2 \,\alpha - \omega_2^2 \, {\rm sinh}^2 \, 
      \alpha = \mu_1 \lbl{3.20}\\
     && -\omega_1^2\, {\rm sinh}^2 \,\alpha + \omega_2^2\, {\rm cosh}^2 \, \alpha = \mu_2 \lbl{3.21}\\
     && (\omega_1^2-\omega_2^2)\, {\rm cosh}\, \alpha \,{\rm sinh}\, \alpha 
  = -\rho_1 \lbl{3.22} .
\ear
The solution of the latter system is $a=\om_1^2$, $b=\om_2^2$, where $\om_1^2$,
$\om_2^2$ are given by Eq.\,(\ref{3.17}).

The Lagrangian and the Hamiltonian in the new coordinates are now\footnote{
As in Case I, also in Case II when $\om_1^2=\om_2^2$,  the system is degenerate and
is not equivalent to the PU oscillator (\ref{3.1}).}
\be
    L=\mbox{$\frac{1}{2}$}({\dot x}'^2 - {\dot y}'^2) - \mbox{$\frac{1}{2}$}
    (\omega_1^2 x'^2 - \omega_2^2 y'^2),
\lbl{3.23}
\ee
\be
    H=\mbox{$\frac{1}{2}$}({\dot x}'^2 - {\dot y}'^2) + \mbox{$\frac{1}{2}$}
    (\omega_1^2 x'^2 - \omega_2^2 y'^2).
\lbl{3.24}
\ee
The latter system can have either positive or negative energy, depending on which
degrees of freedom, $x'$ or $y'$ are more excited.

By employing a procedure analogous to the one reviewed above,
many authors\,\ci{Mostafazadeh}--\ci{Nucci}
have come to the conclusion that the Pais-Uhlenbeck oscillator satisfying
the equations of motion (\ref{3.1}) can be described in terms of a positive
definite Hamiltonian. Since quantization of such a system is straightforward,
those authors concluded that this resolves ``the ghost problem" of the
PU oscillator.

But in the literature it is widely recognized\,\ci{Woodard} that the
quantization of a {\it non interacting} PU oscillator or any system with indefinite
energy is not problematic, if performed correctly so that the correspondence
principle is satisfied. This means that the classical as well as the quantized system has
both, positive and negative energy states, whereas the norm of the quantum
states is always positive. According to those authors, the problems occurs in the
presence of an interaction that couples the positive and negative energy
degrees of freedom, because then the system becomes unstable. But
some authors have observed that interacting systems with positive and
negative energies can be stable as well.

\section{Illustrative example: The system of two equal frequency oscillators}

\subsection{Uncoupled oscillators}

Let us consider a toy model, described by the following Lagrangian and the
Hamiltonian\footnote{
Though this system, because of the degeneracy $\om_1^2=\om_2^2 = \om^2$ is
not equivalent to the PU oscillator, it serves well for illustration of the main points.
It can be straightforwardly generalized to the case of unequal frequencies
$\om_1^2 \neq \om_2^2$ of the two oscillators.}
\be
    L = \frac{1}{2}(\dot x^2  - \dot y^2 ) 
      - \frac{1}{2}\omega ^2 (x^2  - y^2 ) .
\lbl{3.25}
\ee
\be
     H = p_x \dot x + p_y \dot y - L = \frac{1}{2}(p_x^2  - p_y^2 ) 
      + \frac{{\omega ^2 }}{2}(x^2  - y^2 )
\lbl{3.26}
\ee
The Hamilton equations of motion are
\be
  \dotx = \lbrace x,H \rbrace =\frac{\p H}{\p p_x}= p_x~,~~~~~~
  {\dot y} = \lbrace y,H \rbrace =\frac{\p H}{\p p_y}= -p_y
\lbl{3.27}
\ee
\be
  {\dot p}_x = \lbrace p_x,H \rbrace = - \frac{\p H}{\p x} = - \om^2 x~,~~~~~~
    {\dot p}_y = \lbrace p_y,H \rbrace = - \frac{\p H}{\p y} = \om^2 y
\lbl{3.28}
\ee
where the Poisson brackets are defined as usual,
\be
  \lbrace x,p_x \rbrace = 1~,~~~~~~~~\lbrace y,p_y \rbrace = 1
\lbl{3.29}
\ee

In the quantized theory we have commutators
\be
   [x,p_x] = i~,~~~~~~~~[y,p_y]=i
\lbl{3.30}
\ee
Introducing
\bear
 &&c_x  = \frac{1}{{\sqrt {2} }}(\sqrt \omega  \,x + 
\frac{i}{{\sqrt \omega  }}\,p_x )\:,\quad c_x^\dg   
= \frac{1}{{\sqrt 2 }}(\sqrt {\omega}  \,x 
- \frac{i}{{\sqrt \omega  }}\,p_x ) \lbl{3.31} \\ 
 &&c_y  = \frac{1}{{\sqrt {2} }}(\sqrt {\omega}  \,y 
+ \frac{i}{{\sqrt {\omega}  }}\,p_y )\:,\quad c_y^\dg   
= \frac{1}{{\sqrt {2} }}(\sqrt {\omega}  \,y 
- \frac{i}{{\sqrt {\omega}  }}\,p_y ) \lbl{3.32}
 \ear
we have
\be
 [c_x ,c_x^\dg  ] = 1\:,\qquad [c_y ,c_y^\dg  ] = 1,
 \lbl{3.33a}
\ee 
\be
  [c_x ,c_y ] = 0~,\qquad  [c_x^\dg  ,c_y^\dg  ] = 0.
\lbl{3.33b} 
\ee
The Hamiltonian (\ref{3.26}) can be written as
\be
      H = \omega \,(c_x^\dg c_x + c_x c_x^\dg - c_y^\dg  c_y- c_y c_y^\dg ).
\lbl{3.34}
\ee
Defining the vacuum according to
\be
     c_x |0\rangle = 0~,~~~~~~~c_y |0\rangle =0.
\lbl{3.35}
\ee
the Hamiltonian becomes
\be
      H = \omega \,(c_x^\dg  c_x  - c_y^\dg  c_y ).
\lbl{3.36}
\ee
All states have positive norms, e.g., 
\be
  \langle 0|cc^\dg |0\rangle = \langle 0|[c,c^\dg] |0\rangle = \langle 0|0\rangle  = 1,
\lbl{3.37}
\ee
regardless of whether the subscript of $c$, $c^\dg$ is $x$ or $y$, because both
commutators in Eq.\,(\ref{3.33a}) are equal to 1.

In the coordinate representation the momenta are
\be
  p_x = - i \frac{\p}{\p x}~,~~~~~~~~p_y = - i \frac{\p}{\p y},
\lbl{3.38}
\ee
whereas the vacuum state and its defining equations become
\be
   \langle x,y|0 \rangle = \psi_0 (x,y)
\lbl{3.39}
\ee
\bear
 &&\frac{1}{2}\left( {\sqrt \omega  x + \frac{1}{{\sqrt \omega  }}\,
\frac{\partial }{{\partial x}}} \right)\psi _0 (x,y) = 0 \lbl{3.40}\\ 
 &&\frac{1}{2}\left( {\sqrt \omega  y + \frac{1}{{\sqrt \omega  }}\,
\frac{\partial }{{\partial y}}} \right)\psi _0 (x,y) = 0. \lbl{3.41} 
\ear
The solution is
\be
     \psi_0 = \frac{2 \pi}{\om} {\rm e}^{-\frac{1}{2}\omega (x^2 + y^2)}  ,
\lbl{3.42}
\ee
and it has {\it positive norm}
\be
   \int \psi_0^2 \, \dd x\, \dd y = 1.
\lbl{3.43}
\ee

The vacuum $\psi_o(x,y)$ and the states $\psi(x,y)$ excited by successive
actio of the operators $c_x^\dg$,$c_y^\dg$ on $\psi_0(x,y)$ are not
invariant under the hyperbolic rotations in the $(x,y)$-space, under
which the Lagrangian (\ref{3.25}) and the Hamiltonian (\ref{3.26}) are
invariant. However, even if solutions are not invariant, the theory
is covariant, because the set of possible solutions in a new reference frame
 corresponds to the set of possible solutions in the old reference
 frame. Thus, though in a new frame $S'$ the vacuum (\ref{3.42})
 becomes
\be
  \psi_0 (x',y') = \frac{2 \pi}{\om} {\rm e}^{-\frac{1}{2} \om \left ((x' {\rm cosh}\, \alpha
  - y' {\rm sinh}\, \alpha)^2 + (-x' {\rm sinh}\,\alpha + y' {\rm cosh}\,\alpha)^2 \right )},
\lbl{3.44}
\ee
there also exists the solution
\be
     \psi'_0(x',y') = \frac{2 \pi}{\om} {\rm e}^{-\frac{1}{2}\omega (x'^2 + y'^2)}  ,
\lbl{3.45}
\ee
which has the same form as $\psi_0 (x,y)$ of Eq.\,(\ref{3.42}). The same is true
for all excited states.

Such a model can be straightforwardly generalized to higher dimensional spaces
with signature $(r,s)$. The Lagrangian and the Hamiltonian are
\be
    L = \frac{1}{2}\dot x^a \dot x_a  - \frac{1}{2}\omega ^2 x^a x_a ,
\lbl{3.46}
\ee
\be
     H= \frac{1}{2} p^a p_a  + \frac{1}{2}\omega ^2 x^a x_a ,
\lbl{3.47}
\ee
where indices are lowered and raised with the metric $\eta_{ab}$ and its
inverse $\eta^{ab}$. The momentum is $p_a  = \partial L/\partial \dot x^a  
= \dot x_a  = \eta _{ab} \dot x^b$. Upon quantization we have
\be
    [x^a ,p_b ] = i\delta ^a _b 
\lbl{3.47a}
\ee

The creation/annihilation operators defined according to
\bear
     &&c^a  = \frac{1}{{\sqrt 2 }}\left( {\sqrt \omega  x^a  + 
    \frac{i}{{\sqrt \omega  \,}}p_a } \right) \lbl{3.48}\\ 
     &&{c^a}^\dg   = \frac{1}{{\sqrt 2 }}\left( {\sqrt \omega  x^a  
    - \frac{i}{{\sqrt \omega  \,}}p_a } \right), \lbl{3.49} 
\ear
are a generalization of the operators $c_x$,$c_x^\dg$,$c_y$,$c_y^\dg$ given
in Eqs.\,(\ref{3.31}),(\ref{3.32}). They satisfy the commutation relations
\be
   [c^a,{c^b}^\dg]=\delta^{ab}~,~~~~~[c^a,c^b]=[{c^a}^\dg,{c^b}^\dg]=0.
\lbl{3.50}
\ee
The Hamiltonian is
\be
    H = \frac{1}{2}\,\omega \,(c_a^\dg  c^a  + c^a c_a^\dg  )
    =  \omega \,\left (c_a^\dg  c^a + \frac{r}{2} - \frac{s}{2}\right ).
\lbl{3.51}
\ee
If vacuum is defined as
\be
     c^a |0 \rangle = 0,
\lbl{3.52}
\ee
then the vacuum expectation value of $H$ is
\be
  \langle H \rangle = \frac{\om}{2}(r-s),
\lbl{3.53}
\ee
which vanishes if $r=s$, that is when the signature is neutral.

In the literature an alternative definition of annihilation/creation operators
is usually employed, namely
\bear
 &&a^a  = \frac{1}{2}\left( {\sqrt \omega  \,x^a  
+ \frac{i}{{\sqrt \omega  }}\,p^a } \right) \lbl{3.54}\\ 
 &&{a^a}^\dg   = \frac{1}{2}\left( {\sqrt \omega  \,x^a  
- \frac{i}{{\sqrt \omega  }}\,p^a } \right), \lbl{3.55} 
\ear
that satisfy
\be
    [a^a,a_b^\dg]={\delta^a}_b~, ~~~~~[a^a,{a^b}^\dg]= \eta^{ab}.
\lbl{3.56a}
\ee

There are two possible definitions of vacuum:

{\it Definition} I. This is t he usual definition,
\be
     a^a |0 \rangle = 0.
\lbl{3.56}
\ee
The Hamiltonian
\be
     H = \frac{1}{2}\omega \,(a^{a\dg } a_a  + a_a a^{a\dg } )
     =  \omega \,\left (a^{a\dg } a_a  + \frac{r}{2} + \frac{s}{2}\right ) ,
\lbl{3.57}
\ee
acting on the states created by ${a^a}^\dg$ has always {\it positive
eigenvalues}.  But the states corresponding to negative signature
have {\it negative norms}, and are therefore called {\it ghosts}.

{\it Definition} II. This is the Cangemi-Jackiw-Zwiebach definition\,\ci{Cangemi},
discussed in Refs.\,\ci{PavsicPseudoHarm,Woodard}. We split
the operators into the positive and negative signature parts,
\bear
    a^a = (a^\ba,a_{\ul a})~, ~~~~~~~~&&{\bar a}=1,2,...,r~;\nonumber\\
  &&{\ul a} = r+1,r+2,...,r+s .
\lbl{3.58}
\ear
Then the Hamiltonian operators (\ref{3.57}) which can be written as
\be
    H = \omega \,\,\left (a^{\bar a\dg } a_{\bar a}  
    + a_{\ul a} {a^{\ul a}}^\dg  + \frac{r}{2} - \frac{s}{2}\right ) ,
\lbl{3.59}
\ee
has positive and negative eigenvalues. There are no negative norm
states. If the signature is neutral, $r=s$, then the vacuum energy vanishes.

The quantization of the system (\ref{3.46}) according to Definition II has
the correct classical limit, and satisfies the correspondence principle.
Woodard\,\ci{Woodard} argues that this is the correct quantization, whereas
the quantization according to Definition I is incorrect.

Unfortunately, many authors who have been aware of the Definition II, considered
such a system as problematic anyway, because in physically realistic situations
there are interactions between positive and negative energy degrees of
freedom. According to the prevailing opinion,  such a system is necessarily unstable
in the presence of interactions. But it has turned out that this is not true. 
Behaviour of the system (\ref{3.25}) in the presence of various interactions has
been studied in Ref.\,\ci{PavsicFirenze,PavsicRhodes}, where it was found
that for suitable interactions the system is stable.

\setlength{\unitlength}{.8mm}

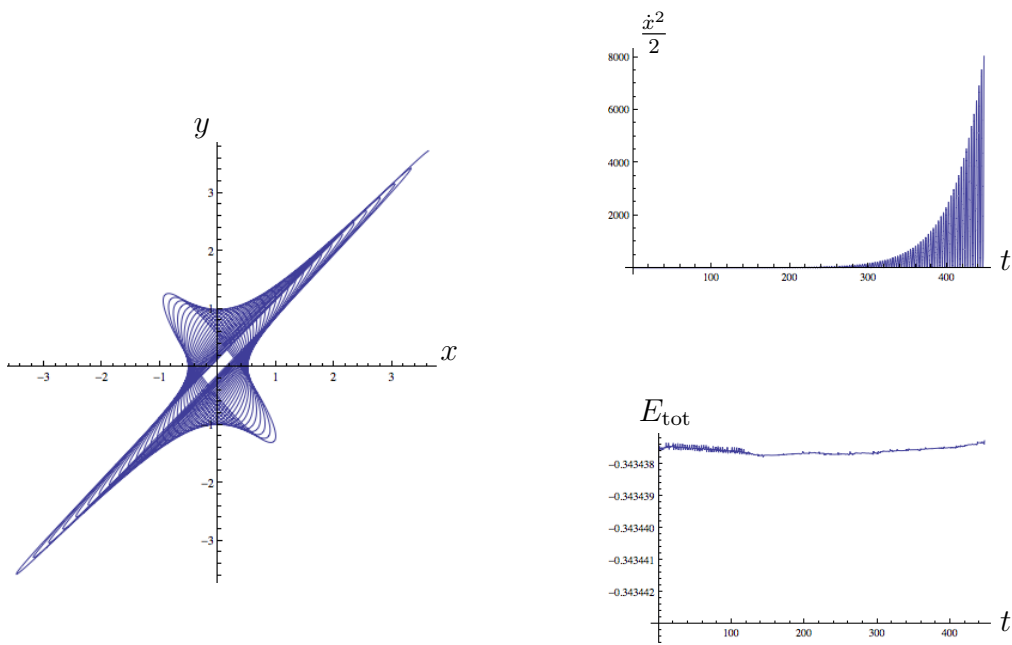
\begin{figure}[h!]
\hs{3mm} \begin{picture}(120,103)(15,0)
\put(20,10){\includegraphics[scale=0.45]{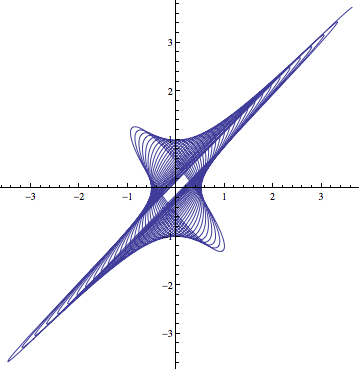}}
\put(120,60){\includegraphics[scale=0.4]{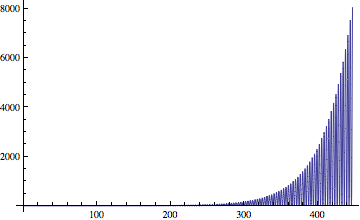}}
\put(120,0){\includegraphics[scale=0.4]{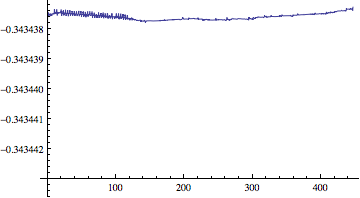}}

\put(92,47){$x$}
\put(51,85){$y$}

\put(125,100){$\frac{{\dot x}^2}{2}$}
\put(185,62){$t$}

\put(125,37){$E_{\rm tot}$}
\put(185,2){$t$}

\end{picture}

\caption{\footnotesize The solution of the system
described by eqs.\, (\ref{4.2}),(\ref{4.3}), for the initial conditions
${\dot x} (0)=1$, ${\dot y} (0)=-1.2$, $x(0)=0$, $y(0)=0.5$.}
\end{figure} 


\subsection{Inclusion of interactions}

Let us now include into the Lagrangian (\ref{3.25}) and additional term
$V_1(x,y)$ that couples the positive energy degree of freedom $x$
and the negative energy degree of freedom $y$, and also include masses
$m_1$, $m_2$:
\be
   L = \frac{1}{2}(m_1 \dot x^2  - m_2 \dot y^2 ) - V\,,\,\,\,\,\,\,\,\,\,\,\,
   V = \frac{\omega }{2}(x^2  - y^2 ) + V_1 .
\lbl{4.1}
\ee
The equations of motion are then
\bear
 &&m_1 \ddot x + \omega ^2 x + \frac{{\partial V_1 }}{{\partial x}} = 0, \lbl{4.2}\\ 
 &&m_2 \ddot y + \omega ^2 y - \frac{{\partial V_1 }}{{\partial y}} = 0. \lbl{4.3} 
\ear

As an example let us first take
\be
   V_1  = \frac{\lambda }{4}(x^2  - y^2 )^2 .
\lbl{4.4}
\ee
and consider two cases:

a) $m_1=m_2 $

In Fig.\,1 are shown the results\,\ci{PavsicFirenze} of numerical solutions\footnote{
Calculations were executed with Mathematica by using NDSolve.}
for $m_1=m_2$. We see that the trajectory in the $(x,y)$-space and
the kinetic energy run into infinity. Such a system is thus unstable.

b) $m_1 \neq m_2$

\setlength{\unitlength}{.8mm}

\begin{figure}[h!]
\hs{3mm}
\begin{picture}(120,115)(25,0)

\put(25,61){\includegraphics[scale=0.3]{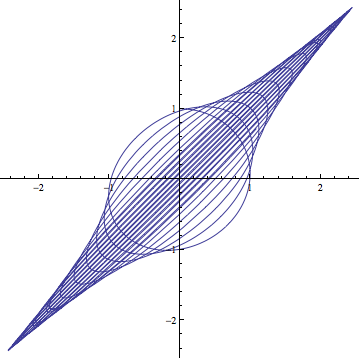}}
\put(90,61){\includegraphics[scale=0.3]{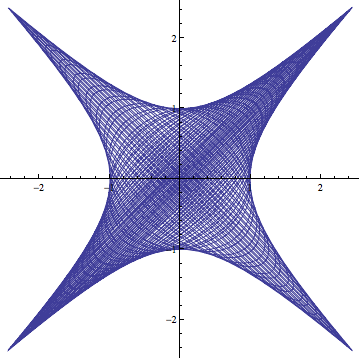}}
\put(150,61){\includegraphics[scale=0.3]{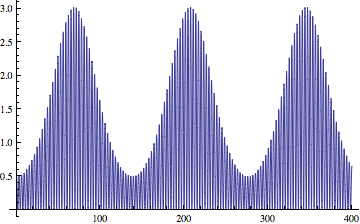}}
\put(25,0){\includegraphics[scale=0.3]{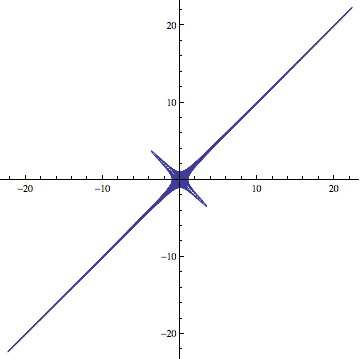}}
\put(90,0){\includegraphics[scale=0.3]{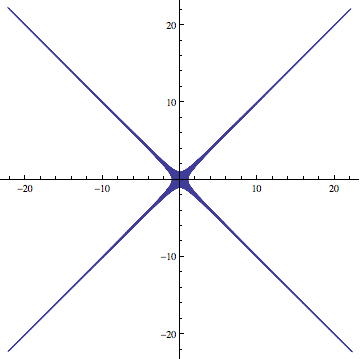}}
\put(150,0){\includegraphics[scale=0.3]{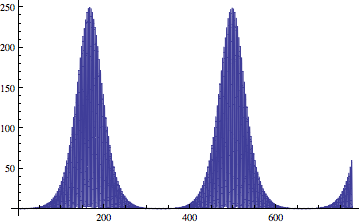}}

\put(73,84){$x$}
\put(46,110){$y$}

\put(139,84){$x$}
\put(112,110){$y$}

\put(139,23){$x$}
\put(111,48){$y$}

\put(73,23){$x$}
\put(46,48){$y$}

\put(149,94){$\frac{{\dot x}^2}{2}$}
\put(199,61){$t$}

\put(149,33){$\frac{{\dot x}^2}{2}$}
\put(200,2){$t$}

\end{picture}

\caption{\footnotesize Up left and middle: The $(x,y)$ plot of the solution to
Eqs.\,(\ref{4.2})(\ref{4.3}) for constants
$m_1=1/1.01$ and $m_2=1$. Up right: The kinetic energy ${\dot x}^2/2$ as
function of time. Low left and middle: The $(x,y)$ plot of the solution
to Eqs.\,(\ref{4.2})(\ref{4.3}) for 
$m_1=1/1.0001$ and $m_2=1$. Low right: The kinetic energy ${\dot x}^2/2$ as
function of time.}
\end{figure}
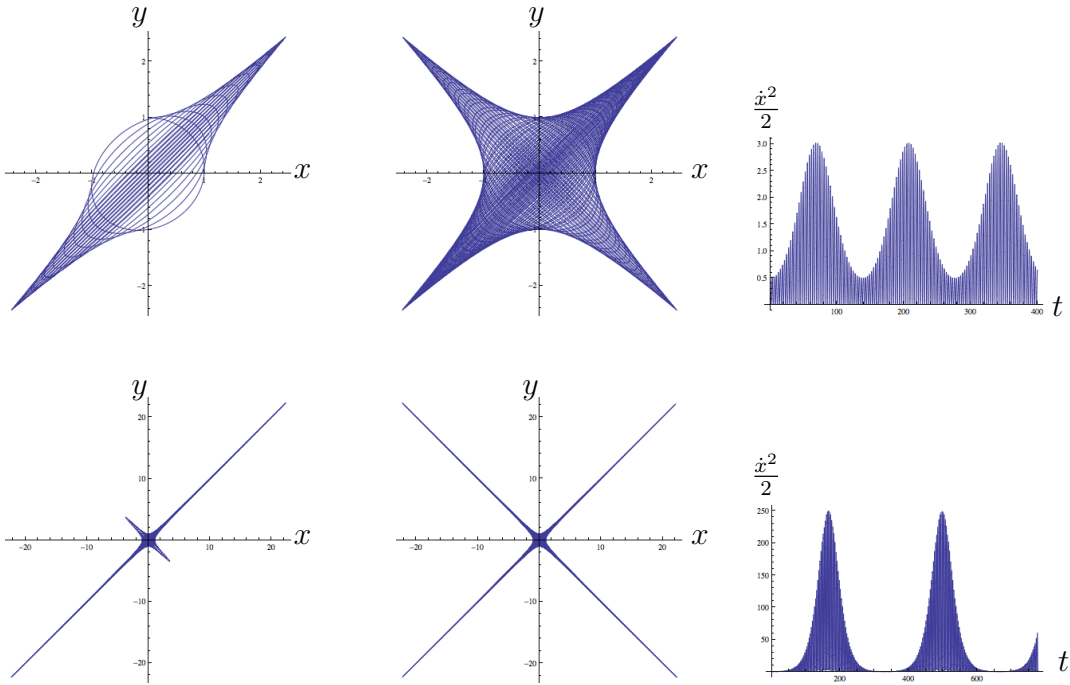 

If masses are different, then something peculiar happens. As shown
in Fig.\,2, the oscillations of the trajectory in the $(x,y)$-space first
increase along one arm\,\ci{PavsicFirenze}. After some time, the second arm forms. The
trajectory remains confined within these two arms. The  envelop of the
kinetic energy oscillations does not run into infinity, but forms the
peaks. If the masses differ very slightly, then the peaks are nearly
separated (Fig.\,2). We see that the system is stable. We checked this
numerically for various initial conditions and positive coupling constants
$\lambda$.
\setlength{\unitlength}{.8mm}

\begin{figure}[h!]
\hs{3mm} \begin{picture}(120,55)(-50,5)

\put(0,0){\includegraphics[scale=0.66]{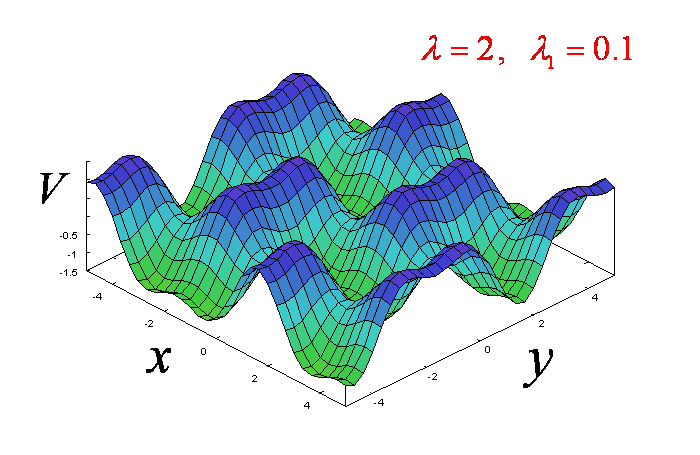}}
\end{picture}

\caption{\footnotesize An example of the potential that is bounded from below and from above (see Eq.\,(\ref{4.5})).}
\end{figure} 
So far we have considered the unbounded potential that was of the form
(\ref{4.1}),(\ref{4.4}). But a realistic potential is bounded from below and from
above. In nature there are
\setlength{\unitlength}{.8mm}

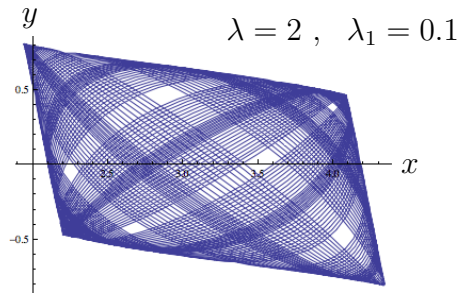
\begin{figure}[h!]
\hs{3mm} \begin{picture}(120,55)(-50,0)

\put(0,0){\includegraphics[scale=0.4]{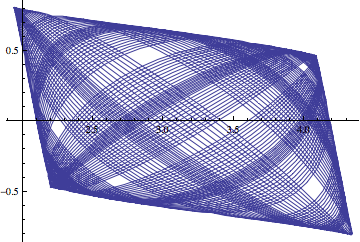}}

\put(65,20){$x$}
\put(2,45){$y$}
\put(36,42){$\lambda=2~,~~\lambda_1=0.1$}

\end{picture}

\caption{\footnotesize The trajectory in the $(x,y)$-space of the solution of the system
described by Eqs.\, (\ref{4.6}),(\ref{4.7}), for the initial conditions
${\dot x} (0)=0.8$, ${\dot y} (0)=0.1$, $x(0)=2.50$, $y(0)=-0.50$.}
\end{figure} 
\nnn no potentials that go into infinity. As an example,
the potential of the form (Fig.\,3)
\be
   V = \frac{\lambda}{4} ({\rm sin}^2 x - {\rm sin}^2 y
    + \lambda_1\, {\rm sin}\,x \, {\rm sin} \,y)
\lbl{4.5}
\ee
was studied in Ref.\,\ci{PavsicRhodes}. The equations of motion are then
\be
  {\ddot x} + \frac{\lambda}{2} ( 2\, {\rm sin}\,x\, {\rm cos}\,x +
  \lambda_1 \, {\rm cos}\,x\, {\rm sin}\,y) = 0,
\lbl{4.6}
\ee
\be
  {\ddot y} + \frac{\lambda}{2} ( 2\, {\rm sin}\,y\, {\rm cos}\,y -
  \lambda_1 \, {\rm sin}\,x\, {\rm cos}\,y) = 0,
\lbl{4.7}
\ee

The results of calculations show that the system, as expected, is stable. If the
initial speed is not to high, the trajectory remains confined within a finite
region of the $(x,y)$-space (Fig.\,4). In the case of an increased initial speed,
the position is not confined, the trajectory can run into infinity (Fig.\,5),
whereas the velocity remains finite. The system behaves as a quasi free
particle subjected to perturbative forces arising from the potential (\ref{4.5}).
Such a system is also stable in the sense that its speed and energy do not
increase into infinity.

\setlength{\unitlength}{.8mm}

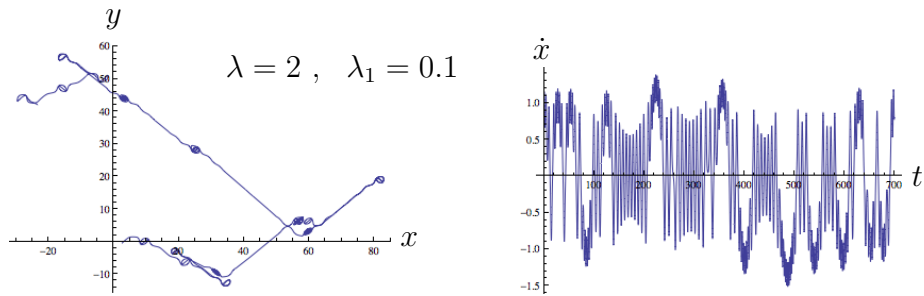
\begin{figure}[h!]
\hs{3mm} \begin{picture}(120,50)(-14,0)

\put(0,0){\includegraphics[scale=0.4]{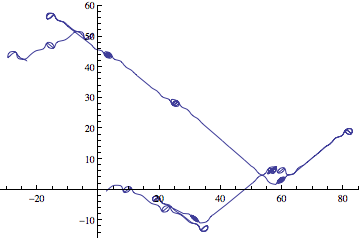}}
\put(85,0){\includegraphics[scale=0.4]{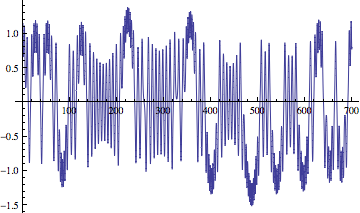}}

\put(65,8){$x$}
\put(16,45){$y$}
\put(36,36){$\lambda=2~,~~\lambda_1=0.1$}

\put(150,18){$t$}
\put(87,39){${\dot x}$}

\end{picture}

\caption{\footnotesize  
The solution of the system described by Eqs.\,(\ref{4.6}),(\ref{4.7}), for the initial conditions
${\dot x} (0)=1$, ${\dot y} (0)=0.9$, $x(0)=2.50$, $y(0)=-0.505$. Left: the trajectory in the $(x,y)$-space. Right: the velocity ${\dot x}$ as function of time $t$.}
\end{figure}

These results show that the systems with a bounded potential do not
exhibit any unusual behaviour even if certain degrees of freedom have
negative energies. There are no runaway solutions and energies do not
increase (or decrease) beyond the boundaries determined by the potential.
Upon correct quantization, such systems are expected to be stable as well
because of the correspondence principle.

\section{Pais-Uhlenbeck oscillator in the presence of quartic self-interaction}

After having discussed in Sec.\,4.2 a toy model interacting system with the
positive energy degree of freedom $x$ and a negative energy degree of freedom
$y$, let us now return to a similar system, which should be equivalent to an
interacting PU oscillator. We will again employ the notation $x'$,$y'$. As discussed
in Sec.\,3, many authors have observed that the {\it non interacting} PU oscillator
can be described either by a Lagrangian which leads to a positive definite
Hamiltonian, or it can be described by a Lagrangian which gives indefinite
Hamiltonian. In Ref.\,\ci{PavsicPUstable} it was shown that in the presence
of an interaction, the system with positive definite Hamiltonian is not equivalent
to a PU oscillator.

\  i) If we start from the Lagrangian
\be
    L=\mbox{$\frac{1}{2}$}({\dot x}'^2 + {\dot y}'^2) - \mbox{$\frac{1}{2}$}
    (\omega_1^2 x'^2 + \omega_2^2 y'^2) - \frac{\lambda}{4} (x'+y')^4,
 \lbl{5.1}
\ee
whose equations of motion are
\be
   {\ddot x'} +\omega_1^2 x' + \lambda (x'+y')^3 = 0,
\lbl{5.2}
\ee
\be
   {\ddot y'} +\omega_2^2 y'+ \lambda (x'+y')^3  = 0,
\lbl{5.3}
\ee
and introduce the new variables
\be
    u=\frac{x'+y'}{\sqrt{2}}~,~~~~~~~~~~v=\frac{x'-y'}{\sqrt{2}} ,
\lbl{5.4}
\ee
we obtain
\be
   L=\mbox{$\frac{1}{2}$}({\dot u}^2 + {\dot v}^2) - \mbox{$\frac{1}{4}$}
    [(\omega_1^2 + \omega_2^2)(u^2+v^2) + 2 (\omega_1^2 - \omega_2^2)u v]
     - \lambda u^4
 \lbl{5.5}
\ee
Introducing
\be
    \mu_1 = \mu_2 = \mbox{$\frac{1}{2}$}(\omega_1^2 + \omega_2^2) ~,~~~~~
     -\rho_1 = \mbox{$\frac{1}{2}$}(\omega_1^2 - \omega_2^2),
\lbl{5.6}
\ee
the equation of motion derived from (\ref{5.5}) are
\be
     {\ddot u} + \mu_1 u -\rho_1 v + 4 \lambda u^3 = 0
\lbl{5.7}
\ee
\be
     {\ddot v} + \mu_2 v -\rho_1 u = 0 ,
\lbl{5.8}
\ee
which, after elimination of $u$, becomes
\be
     v^{(4)} + (\mu_1 + \mu_2) {\ddot v} + (\mu_1 \mu_2 - \rho_1^2) v +
   4 \lambda \rho_1 ({\ddot v} + \mu_2 v)^3 =0 ,
\lbl{5.9}
\ee
The interaction term is nonlinear in the second order derivative ${\ddot v}$.

In the system (\ref{5.7}),(\ref{5.8}) we may as well eliminate $v$. Then we obtain
the equation
\be
     u^{(4)} + (\mu_1 + \mu_2) {\ddot u} + (\mu_1 \mu_2 - \rho_1^2) u +
   4 \mu_2 \lambda u^3 + 4 \lambda \frac{\dd^2}{\dd t^2} \left ( u^3 \right ) = 0,
\lbl{5.10}
\ee
which also contains a nonlinear term.

Neither Eq.\,(\ref{5.9}) nor (\ref{5.10}) does contain a simple interaction term
that could be derived from a potential. This reveals that the system described
by the Lagrangian (\ref{5.1}) is not equivalent to a self-interacting PU
oscillator.

(ii) The situation is different if we start from  the Lagrangian
\be
    L=\mbox{$\frac{1}{2}$}({\dot x}'^2 - {\dot y}'^2) - \mbox{$\frac{1}{2}$}
    (\omega_1^2 x'^2 - \omega_2^2 y'^2) - \frac{\lambda}{4} (x'+y')^4 ,
 \lbl{5.11}
\ee
which in terms oof the new variables (\ref{5.4}) reads
\be
   L={\dot u} {\dot v} - \mbox{$\frac{1}{4}$}
    [(\omega_1^2 - \omega_2^2)(u^2+v^2) + 2 (\omega_1^2 + \omega_2^2)u v]
     - \lambda u^4 .
 \lbl{5.12}
\ee
The equations of motion are now
\be
     {\ddot u} + \mu_1 u -\rho_1 v = 0
\lbl{5.13}
\ee
\be
     {\ddot v} + \mu_2 v -\rho_1 u + 4 \lambda u^3= 0 ,
\lbl{5.14}
\ee
where $\mu_1$,$\mu_2$ and $\rho_1$ are given in Eq.\,(\ref{5.6}). Eliminating
$v$, we have
\be
    u^{(4)} + (\mu_1 + \mu_2) {\ddot u} +(\mu_1 \mu_2 -\rho_1^2) u +
  4 \rho_1 \lambda u^3 = 0.
\lbl{5.15}
\ee
This can be rewritten in terms of $\om_1^2$,$\om_2^2$ by using Eq.\,(\ref{5.6}):
\be
    u^{(4)} + (\omega_1^2 + \omega_2^2) {\ddot u} +\omega_1 \omega_2^2 u 
    -\Lambda u^3 = 0 ,
\lbl{5.16}
\ee
where $\Lambda=2 (\omega_1^2-\omega_2^2) \lambda$. The corresponding
Lagrangian is
\be
    L=\mbox{$\frac{1}{2}$} \left [ {\ddot u}^2 -(\omega_1^2 + \omega_2^2) 
    {\ddot u}^2 +\omega_1 \omega_2^2 u^2 \right ]
    +\mbox{$\frac{1}{4}$} \Lambda u^4 .
\lbl{5.17}
\ee
This is the Lagrangian for a PU oscillator with a quartic self-interaction term.
The variable $u$ here is in fact the same as the variable $x$ used in
Eqs.\,(\ref{2.7})--(\ref{2.9}) in Sec.\,3.

\setlength{\unitlength}{.8mm}

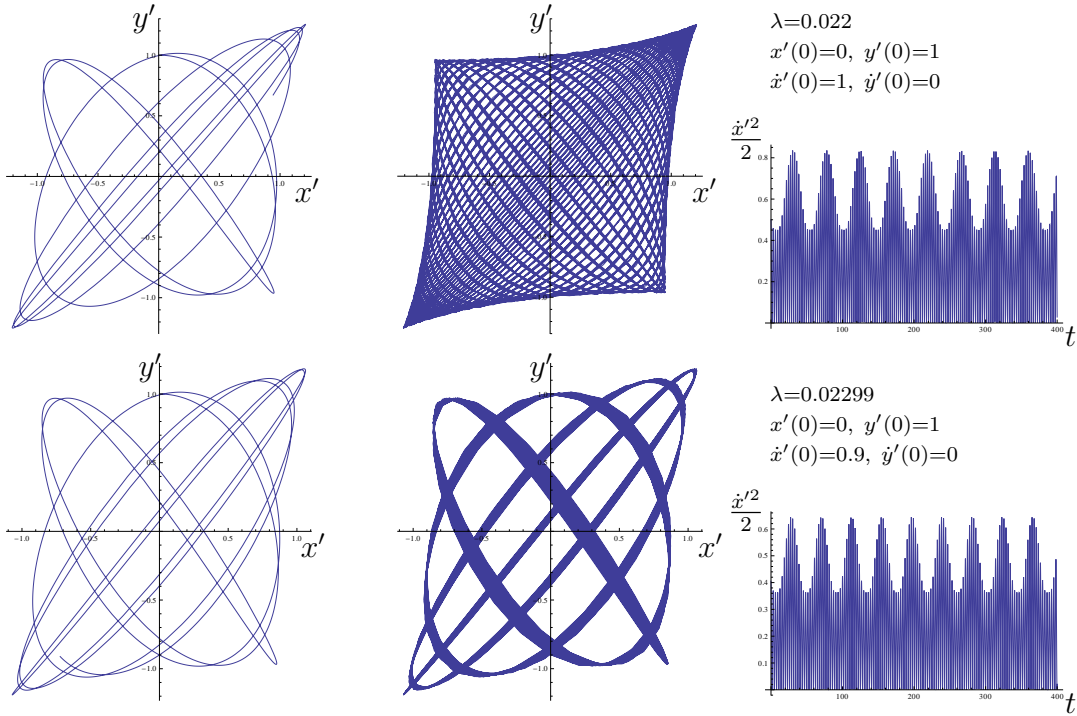
\begin{figure}[h!]
\hs{3mm}
\begin{picture}(120,112)(25,0)

\put(25,61){\includegraphics[scale=0.4]{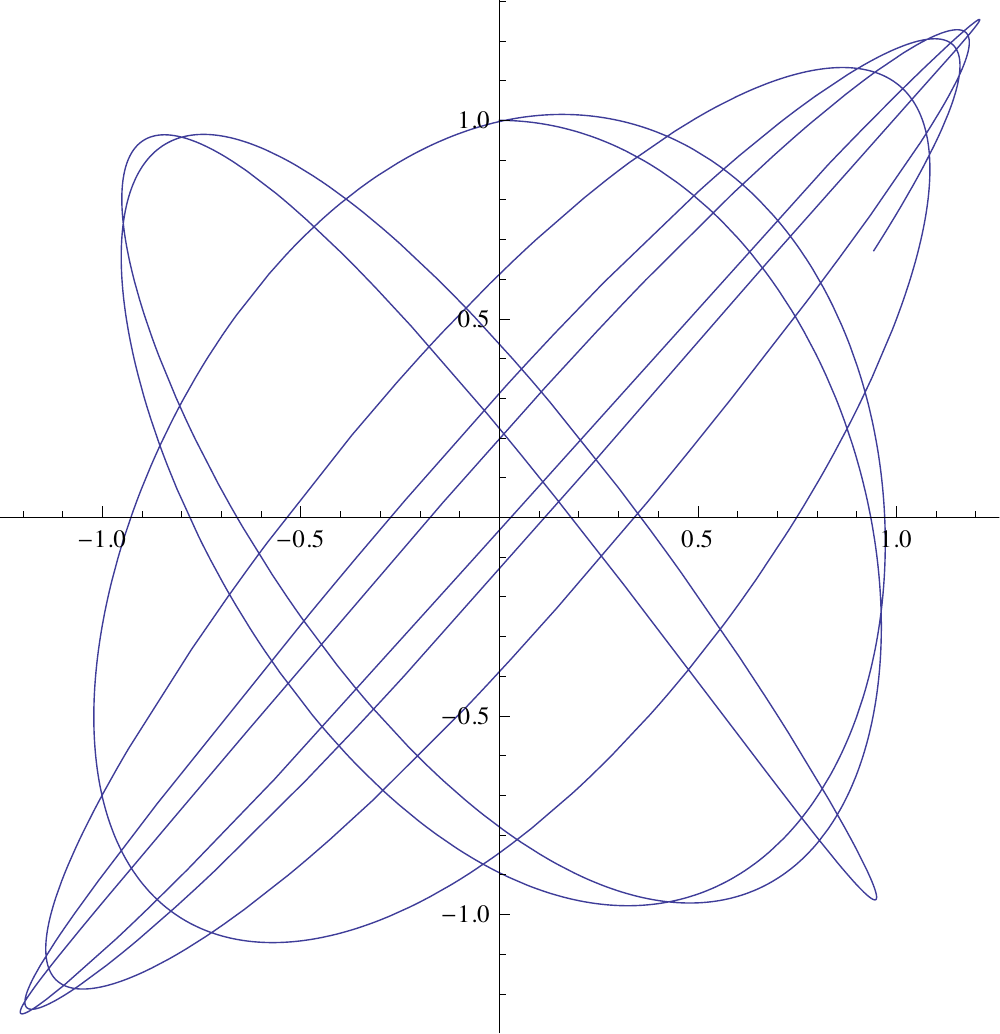}}
\put(90,61){\includegraphics[scale=0.4]{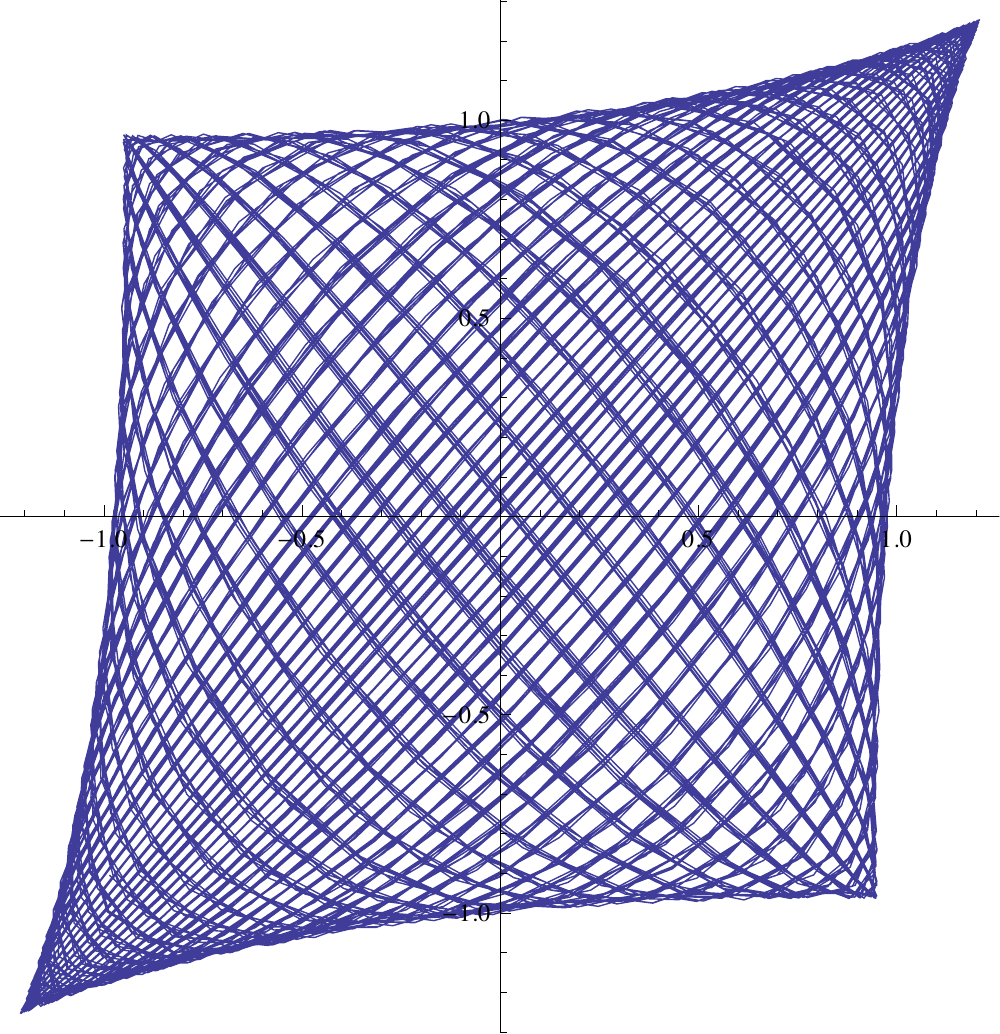}}
\put(150,61){\includegraphics[scale=0.4]{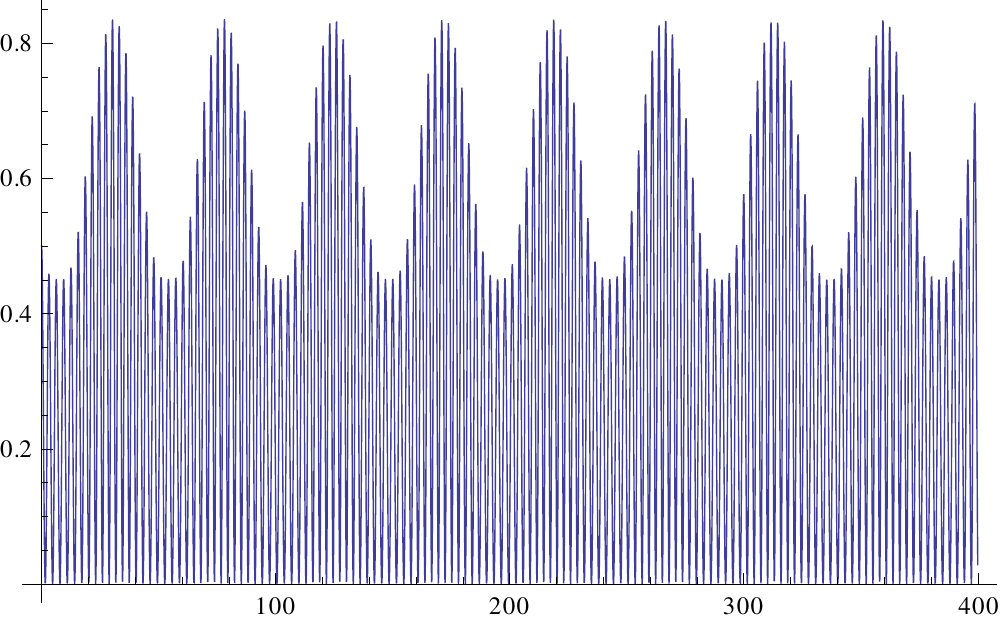}}
\put(25,0){\includegraphics[scale=0.4]{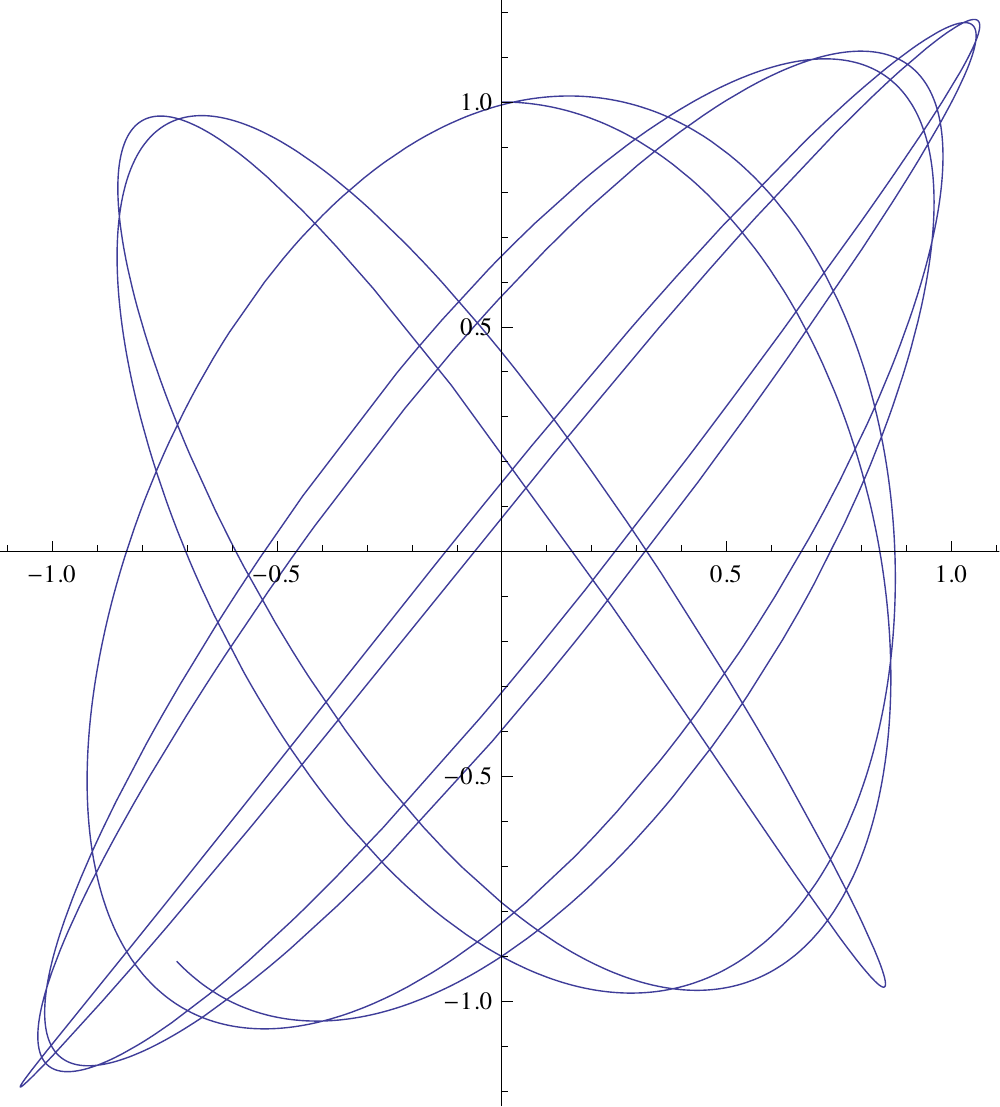}}
\put(90,0){\includegraphics[scale=0.4]{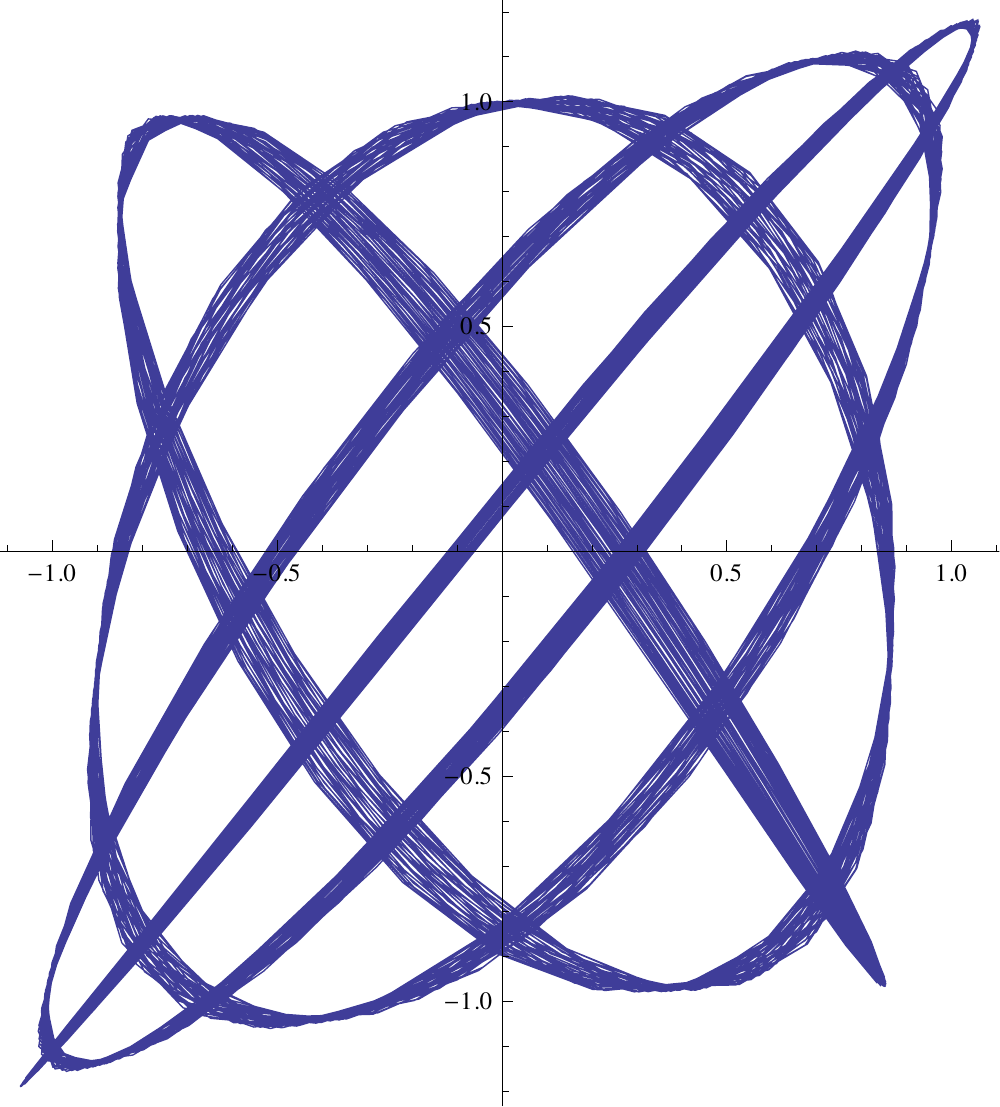}}
\put(150,0){\includegraphics[scale=0.4]{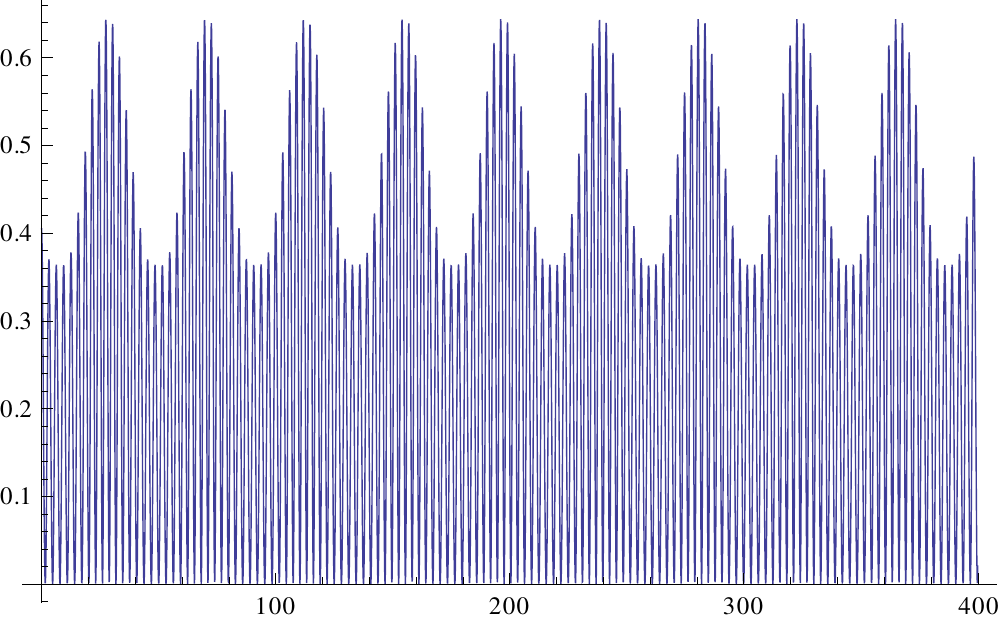}}

\put(73,83){$x'$}
\put(46,112){$y'$}

\put(139,83){$x'$}
\put(112,112){$y'$}

\put(74,24){$x'$}
\put(47,54){$y'$}

\put(140,24){$x'$}
\put(112,54){$y'$}

\put(145,92){$\frac{{\dot x}'^2}{2}$}
\put(201,59){$t$}

\put(145,30){$\frac{{\dot x}'^2}{2}$}
\put(201,-2){$t$}

\put(152,110){$^{\lambda=0.022}$}
\put(152,105){$^{x'(0)=0,~y'(0)=1}$}
\put(152,100){$^{{\dot x}'(0)=1,~{\dot y}'(0)=0}$}
\put(152,48){$^{\lambda=0.02299}$}
\put(152,43){$^{x'(0)=0,~y'(0)=1}$}
\put(152,38){$^{{\dot x}'(0)=0.9,~{\dot y}'(0)=0}$}

\end{picture}

\caption{\footnotesize Solutions of the equations of motion derived from
the Lagrangian  (\ref{5.11} for
different values of the coupling constant $\lambda$ and different
initial conditions.
Left and middle: the trajectories in the $(x',y')$ space.
Right: The kinetic energy ${\dot x}'^2/2$ as function of time. The oscillations
within the envelope are so fine that they fill the diagram.}
\end{figure} 
By applying the Ostrogradski formalism to (\ref{5.17}) we obtain the
Hamiltonian (\ref{2.9}) which is not positive definite. Inclusion of an
interaction term into the free PU Lagrangian reveals that such a second
order system has to be processed \`a la Ostrogradski. Alternative ways
discussed in the litarature\,\ci{Mostafazadeh}--\ci{Nucci}, where the PU oscillator is described 
in terms of positive energies only, are applicable to the free oscillator,
but not to a (self) interacting one.

It used to be taken for granted that a system with a Lagrangian of the
sort (\ref{5.11}), or, equivalently (\ref{5.16}) is unstable. But Smilga\,\ci{Smilga1}
has found that this is not necessarily so. For a certain range of initial
velocity and coupling constant the system (\ref{5.11}) can be stable.
This has been studied in Refs.\,\ci{PavsicPUstable}, where it was found
by numerical calculations that the system was stable if the initial
velocity ${\dot x}'(0)$, ${\dot y'}(0)$ and the coupling constant
$\lambda$ were below certain critical values (Fig.\,6). Above the critical
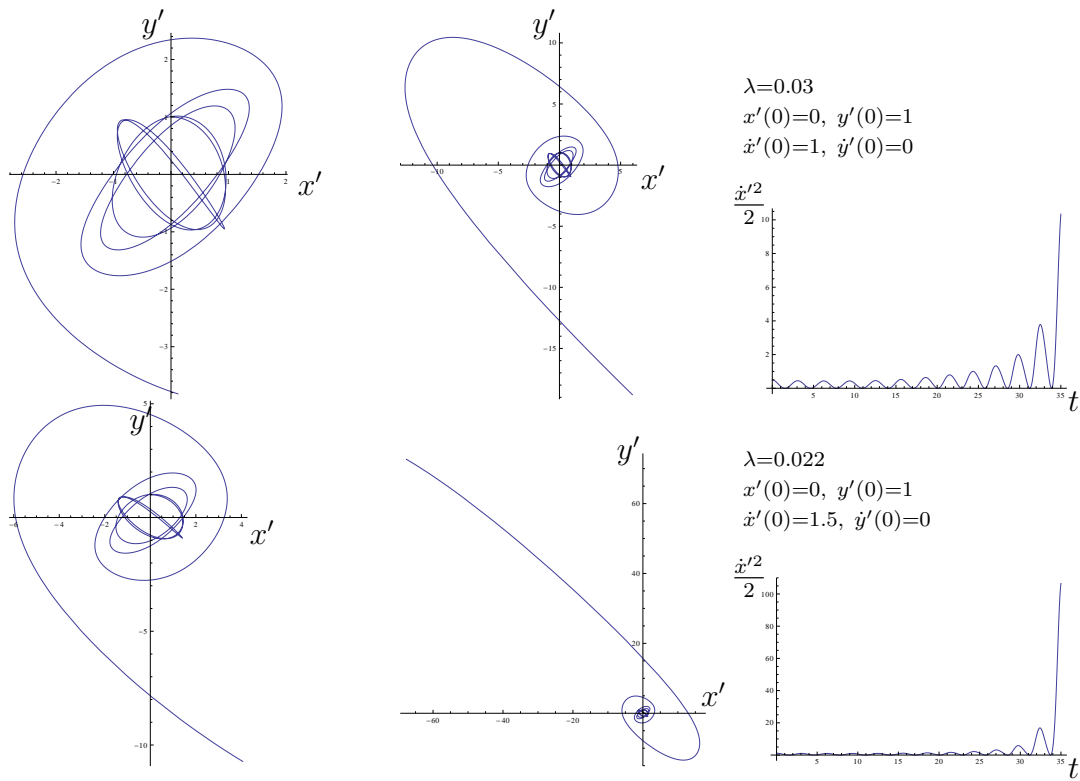
\begin{figure}[h!]
\hs{3mm}
\begin{picture}(120,125)(25,0)

\put(25,61){\includegraphics[scale=0.4]{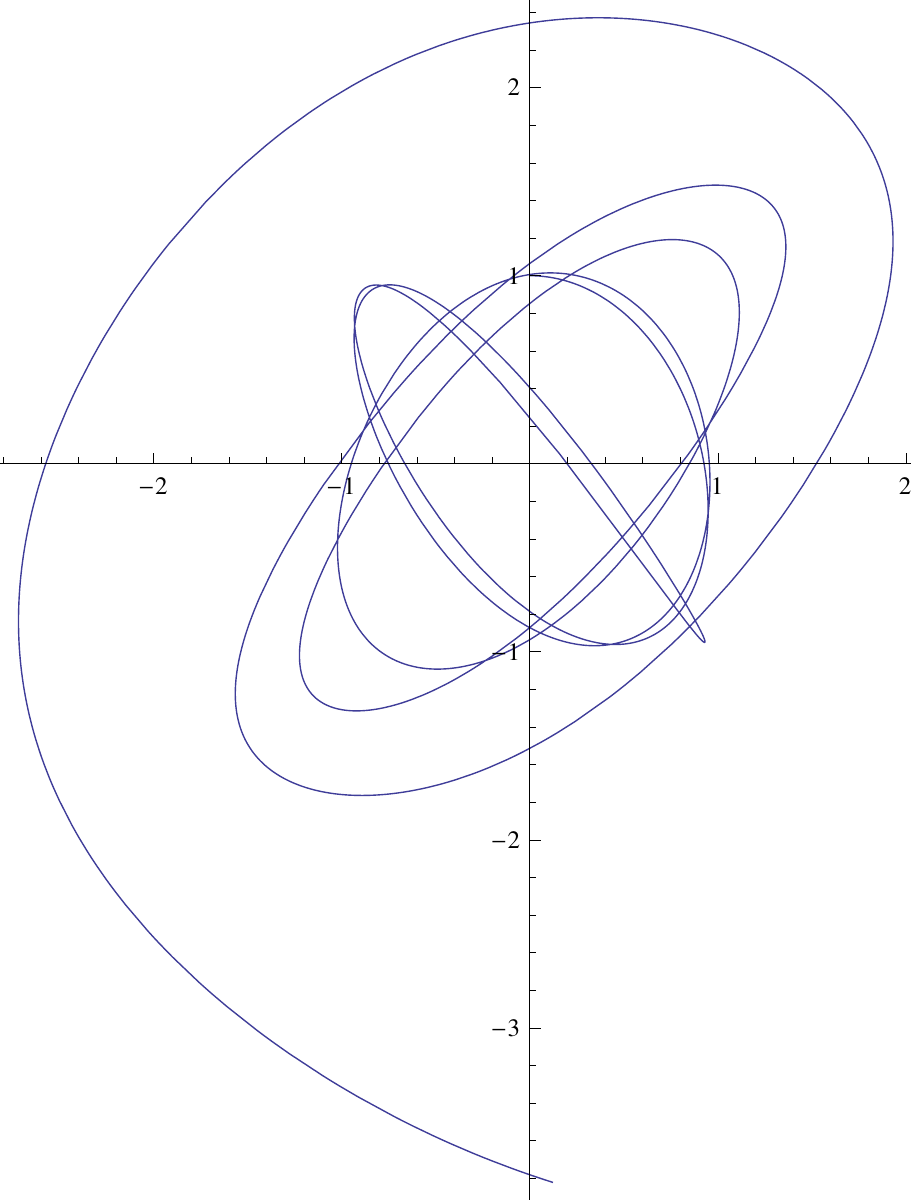}}
\put(90,61){\includegraphics[scale=0.4]{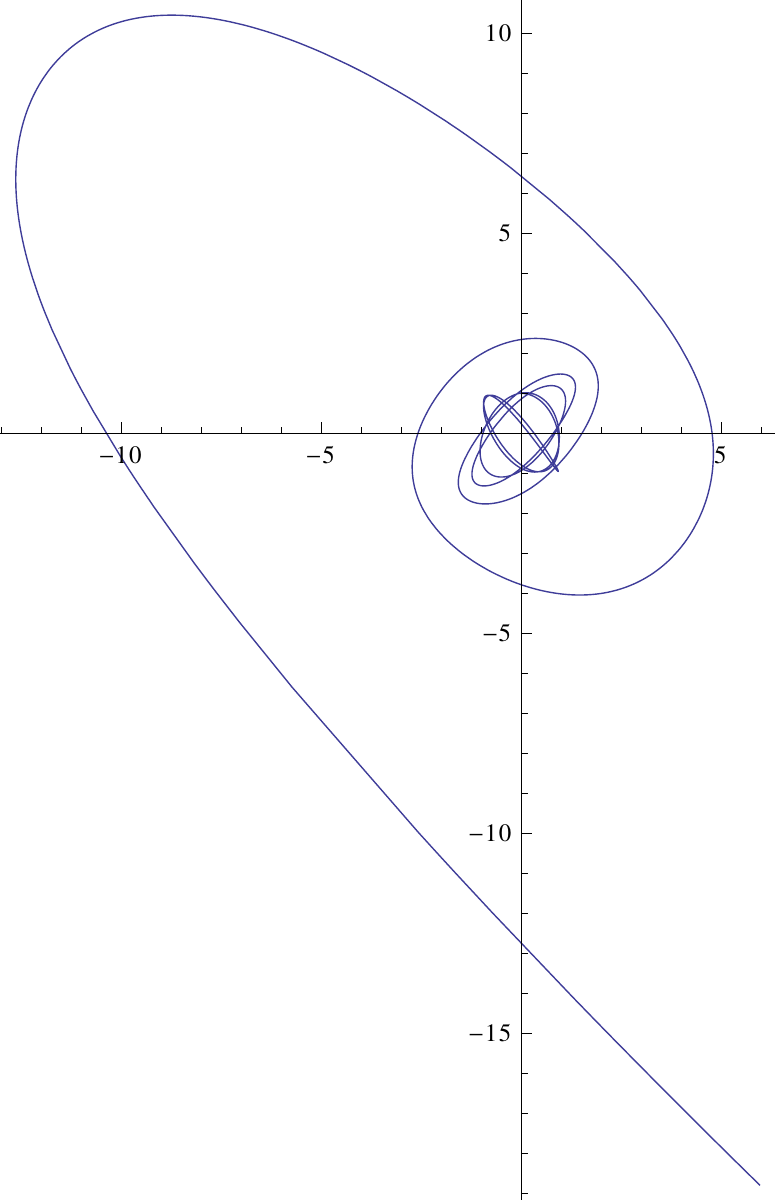}}
\put(150,61){\includegraphics[scale=0.4]{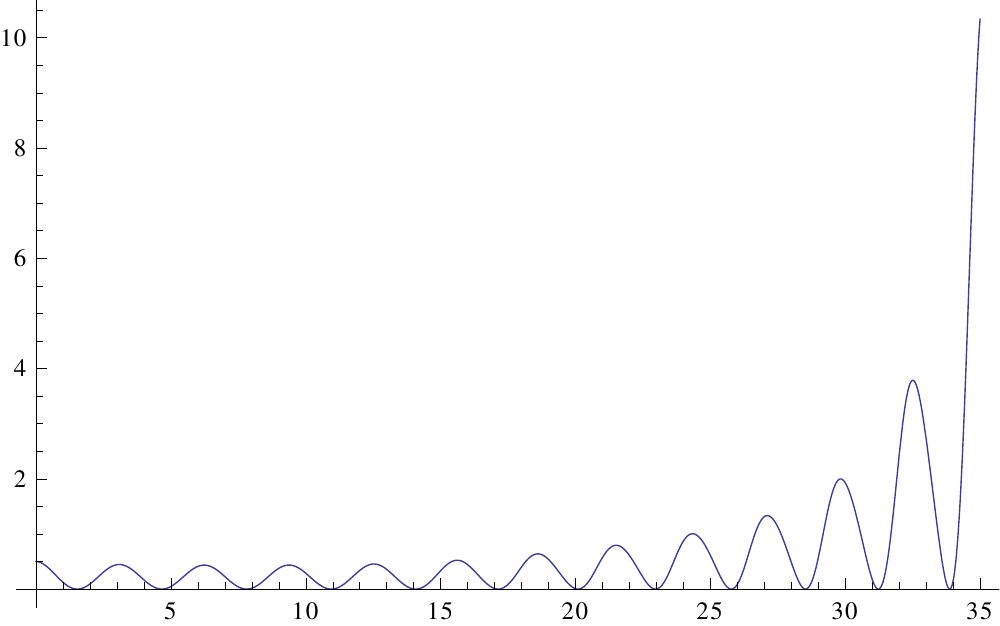}}
\put(25,0){\includegraphics[scale=0.4]{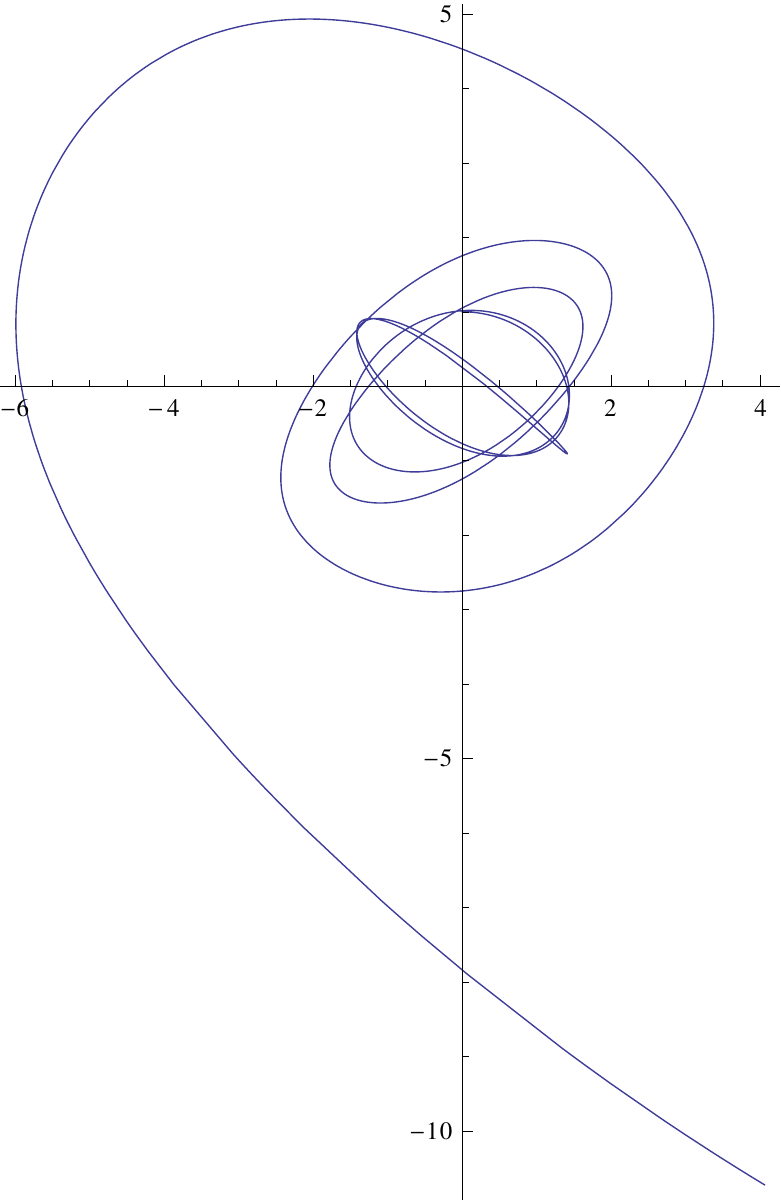}}
\put(90,0){\includegraphics[scale=0.4]{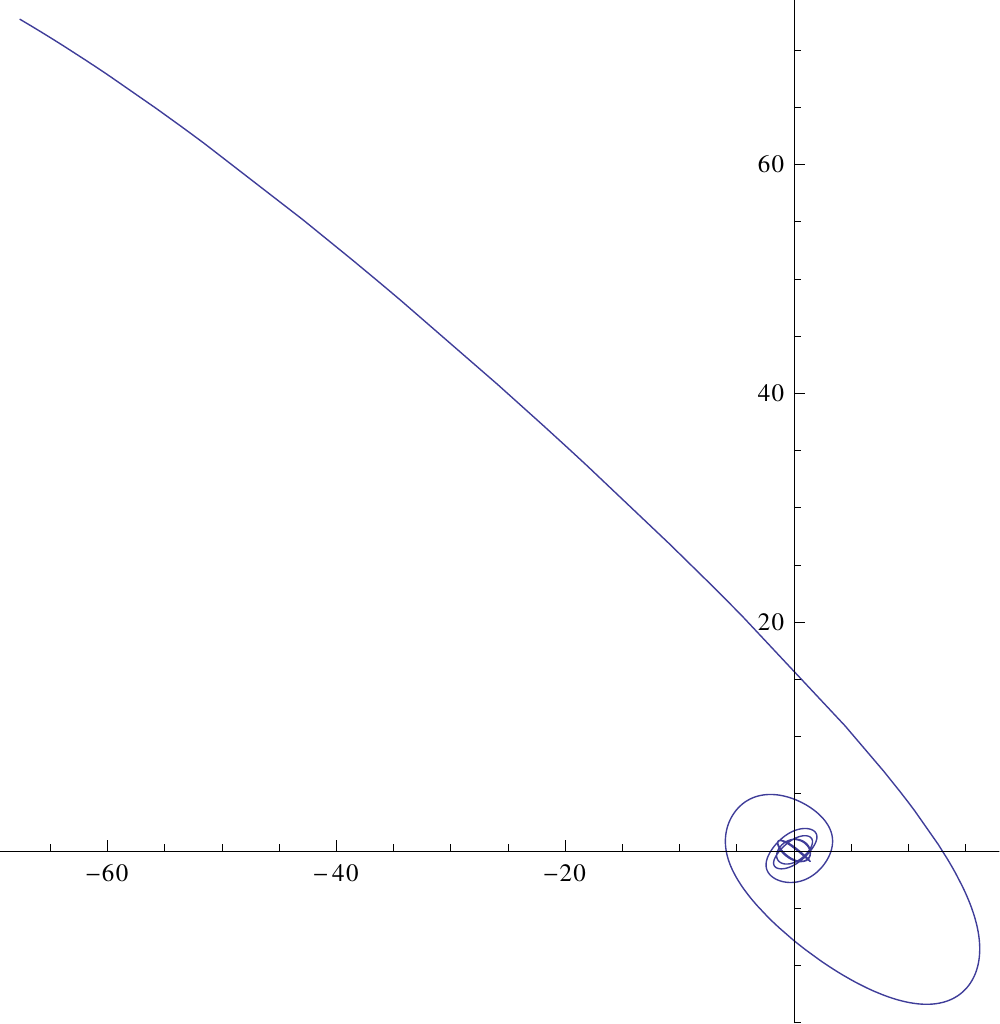}}
\put(150,0){\includegraphics[scale=0.4]{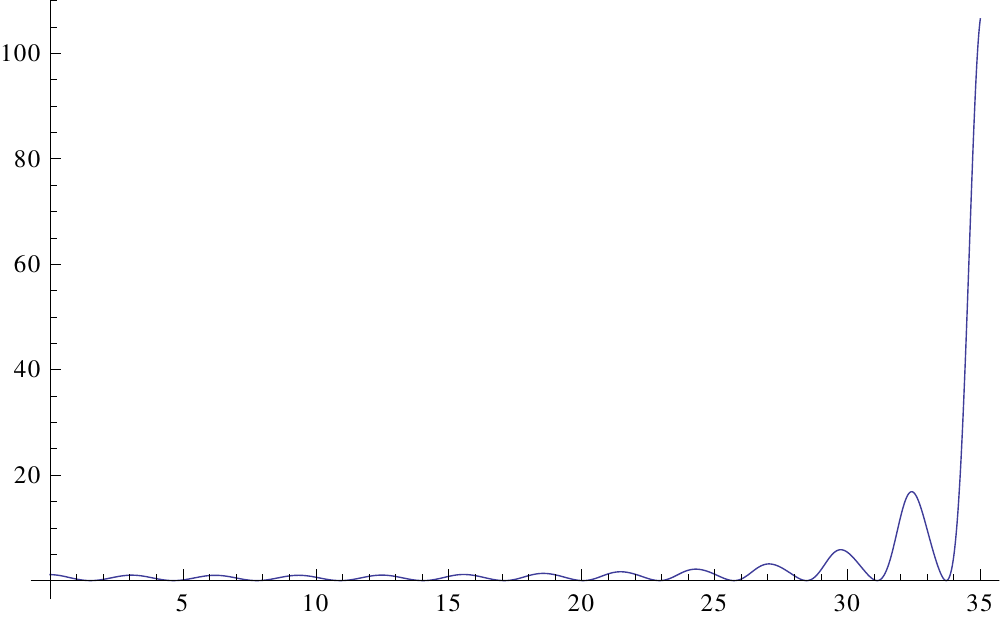}}

\put(73,95){$x'$}
\put(47,122){$y'$}

\put(130,96){$x'$}
\put(112,122){$y'$}

\put(65,37){$x'$}
\put(45,56){$y'$}

\put(140,10){$x'$}
\put(126,51){$y'$}

\put(145,92){$\frac{{\dot x}'^2}{2}$}
\put(201,59){$t$}

\put(145,30){$\frac{{\dot x}'^2}{2}$}
\put(201,-2){$t$}

\put(147,110){$^{\lambda=0.03}$}
\put(147,105){$^{x'(0)=0,~y'(0)=1}$}
\put(147,100){$^{{\dot x}'(0)=1,~{\dot y}'(0)=0}$}
\put(147,48){$^{\lambda=0.022}$}
\put(147,43){$^{x'(0)=0,~y'(0)=1}$}
\put(147,38){$^{{\dot x}'(0)=1.5,~{\dot y}'(0)=0}$}

\end{picture}

\caption{\footnotesize Up: By increasing the $\lambda$, the system becomes
unstable. The trajectory and the kinetic energy escape into infinity.
Down: Similarly, by increasing the initial velocity, the system also becomes
unstable. }
\end{figure} 

\nnn values the system exhibited runaway behaviour (Fig.\,7). Very close to the
critical value of $\lambda$ the solution oscillated seemingly stably within
a confined region of the $(x',y')$-space, but after a long time it escaped
into infinity (Fig.\,8). It is reasonable to expect that a similar behaviour could
occur in certain solid state system that can eventually be described by a
higher order derivative field theory \`a la Eq.\,(\ref{2.1}). This could have
far reaching consequences and applications (see also Ref.\,\ci{PavsicFirenze}).
\begin{figure}[h!]
\hs{3mm}
\begin{picture}(120,60)(25,0)

\put(25,0){\includegraphics[scale=0.4]{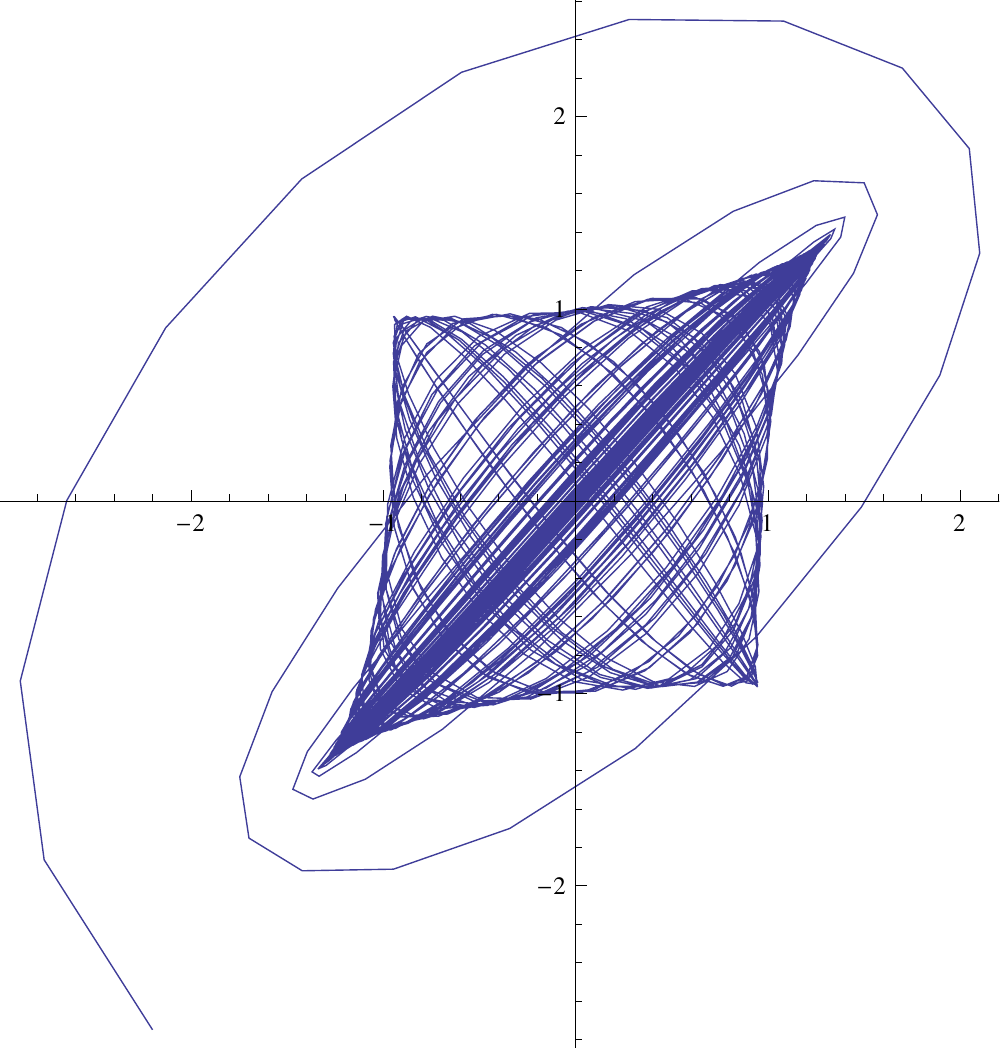}}
\put(90,0){\includegraphics[scale=0.4]{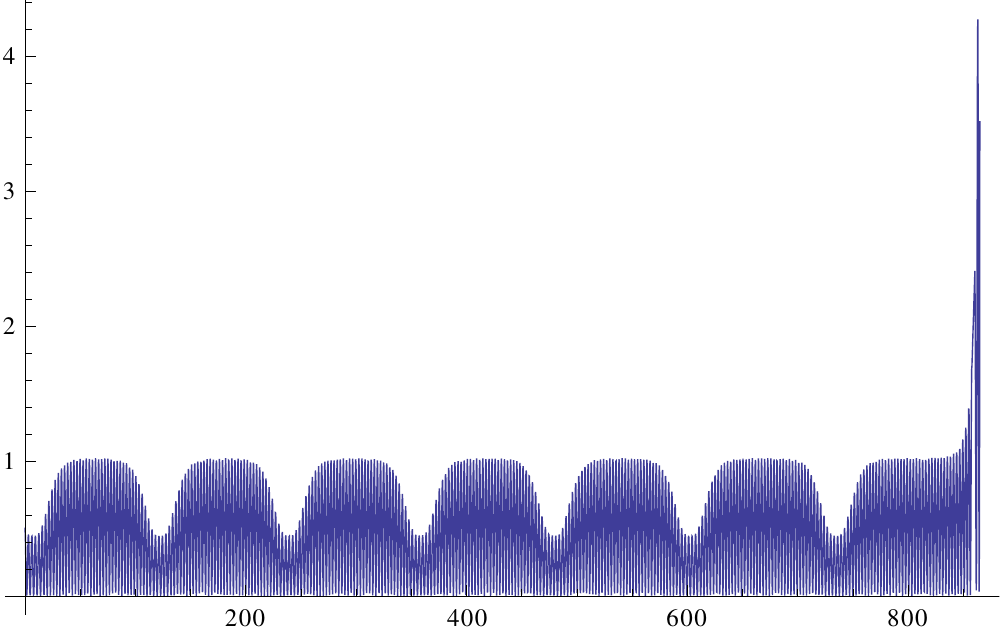}}
\put(150,0){\includegraphics[scale=0.4]{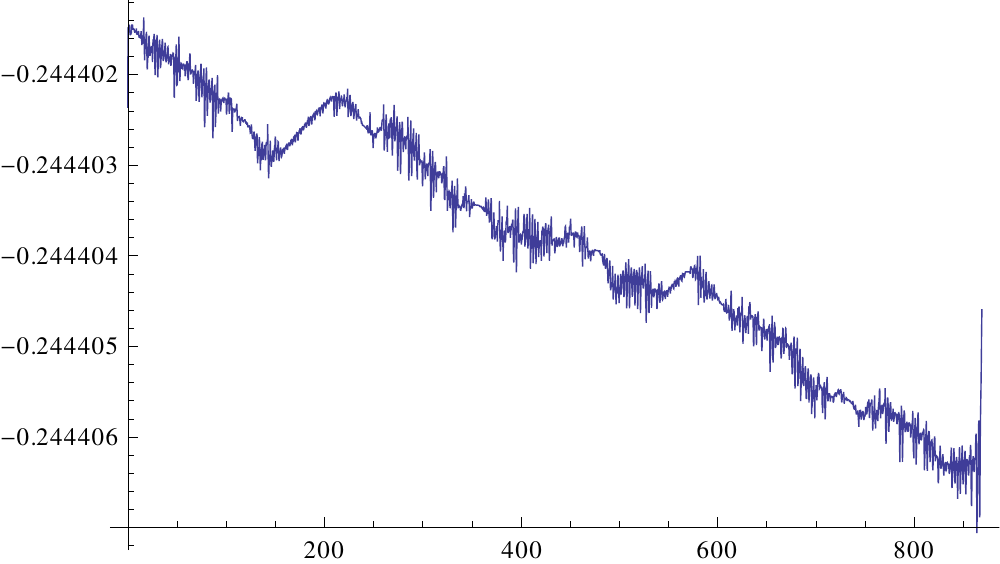}}

\put(74,23){$x$}
\put(51,53){$y$}

\put(85,30){$\frac{{\dot x}^2}{2}$}
\put(141,-2){$t$}

\put(152,30){$E_{\rm tot}$}
\put(201,-2){$t$}

\put(130,48){$^{\lambda=0.02299}$}
\put(130,43){$^{x'(0)=0,~y'(0)=1}$}
\put(130,38){$^{{\dot x}'(0)=0.999851,~{\dot y}'(0)=0}$}

\end{picture}

\caption{\footnotesize At certain values of $\lambda$ and the initial conditions,
the system behaves stably for a long time, before it finally escapes to infinity.
The total energy $E_{\text tot}$ remains constant within the numerical error.
}
\end{figure}
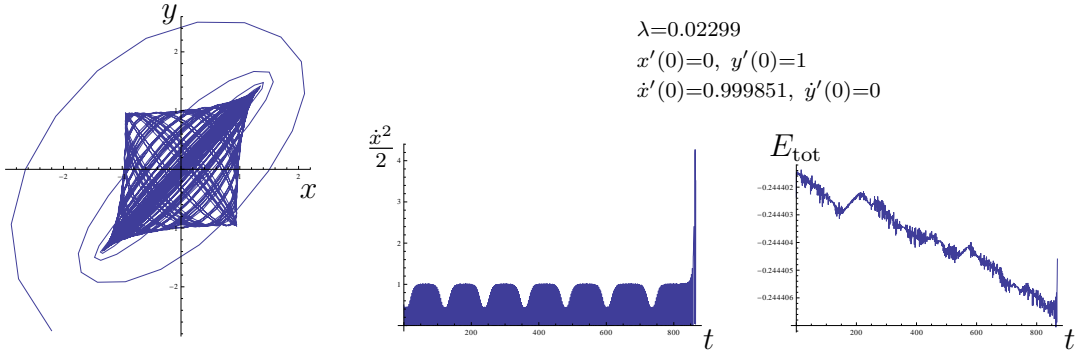

\section{The Cases of Stable Interacting Pais-Uhlenbeck Oscillator}

\  (i) {\it A Bounded Interaction Term}

Let us now investigate how the system behaves if the interaction potential
is bounded from below and form above. As an example the interaction term
$\frac{1}{4}{\rm sin}^4 \,(x'+y')$ was considered in Ref.\,\ci{PavsicPUstable}.
Instead of the Lagrangian (\ref{5.11}) we then have
\be
    L=\mbox{$\frac{1}{2}$}({\dot x}'^2 - {\dot y}'^2) - \mbox{$\frac{1}{2}$}
    (\omega_1^2 x'^2 - \omega_2^2 y'^2) - \frac{\lambda}{4} {\rm sin}^4 (x'+y') ,
 \lbl{6.1}
\ee
which gives the following equations of motion:
\be
   {\ddot x'} +\omega_1^2 x' + \lambda \text{sin}^3\, (x'+y') 
   \text{cos} \,(x'+y')= 0,
\lbl{6.1a}
\ee
\be
   {\ddot y'} +\omega_2^2 y'- \lambda \text{sin}^3\, (x'+y') 
   \text{cos} \,(x'+y')= 0.
\lbl{6.1b}
\ee
By solving the above equation numerically, it was found\,\ci{PavsicPUstable}
that such a system is stable for all initial velocities and for all positive values of the
coupling constant $\lambda$ (see Fig.\,9).

\setlength{\unitlength}{.8mm}

\begin{figure}[h!]
\hs{3mm}
\begin{picture}(120,115)(25,-5)

\put(25,55){\includegraphics[scale=0.4]{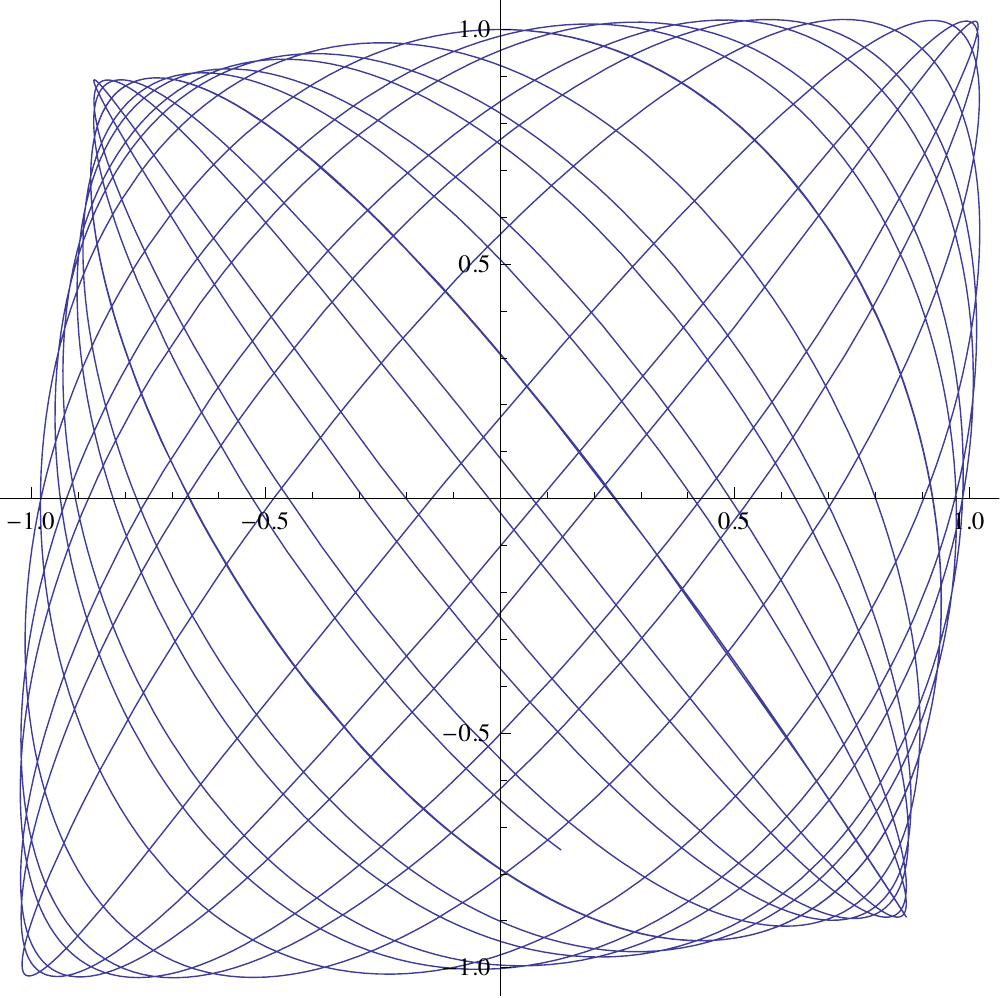}}
\put(90,55){\includegraphics[scale=0.4]{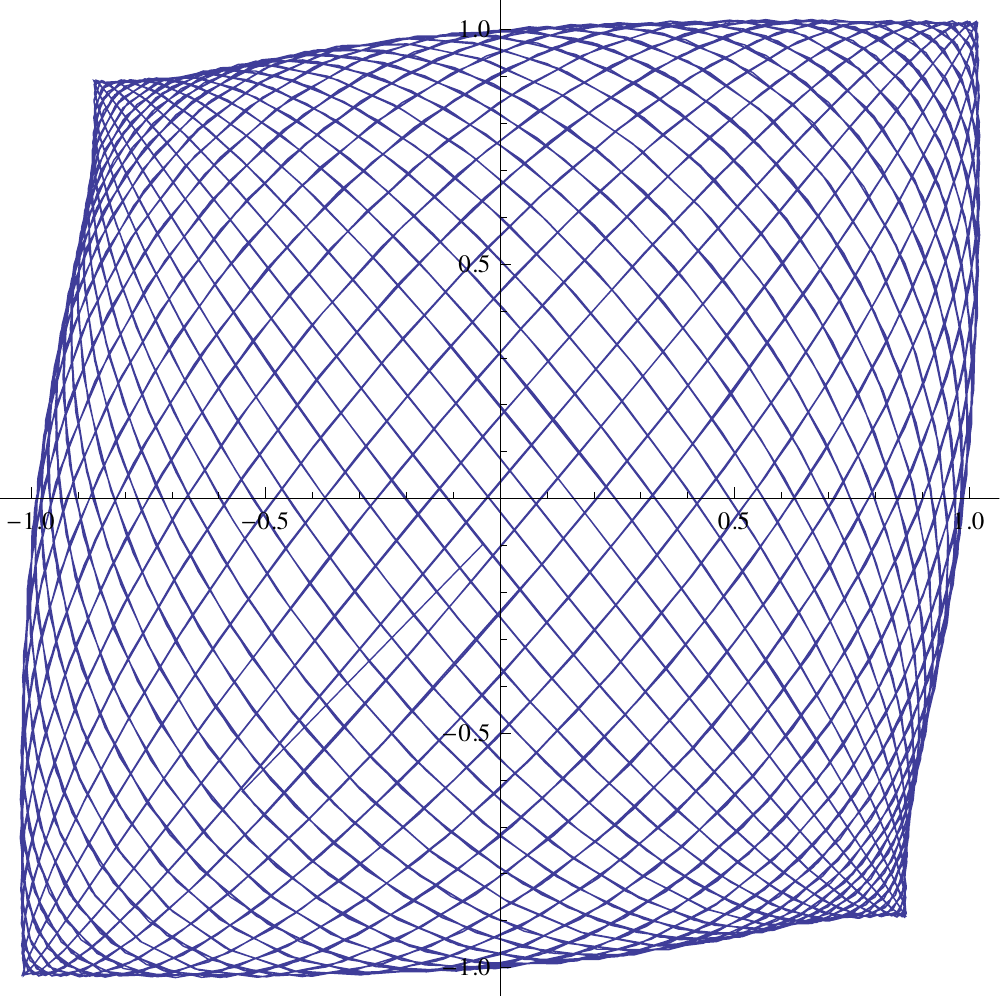}}
\put(150,55){\includegraphics[scale=0.4]{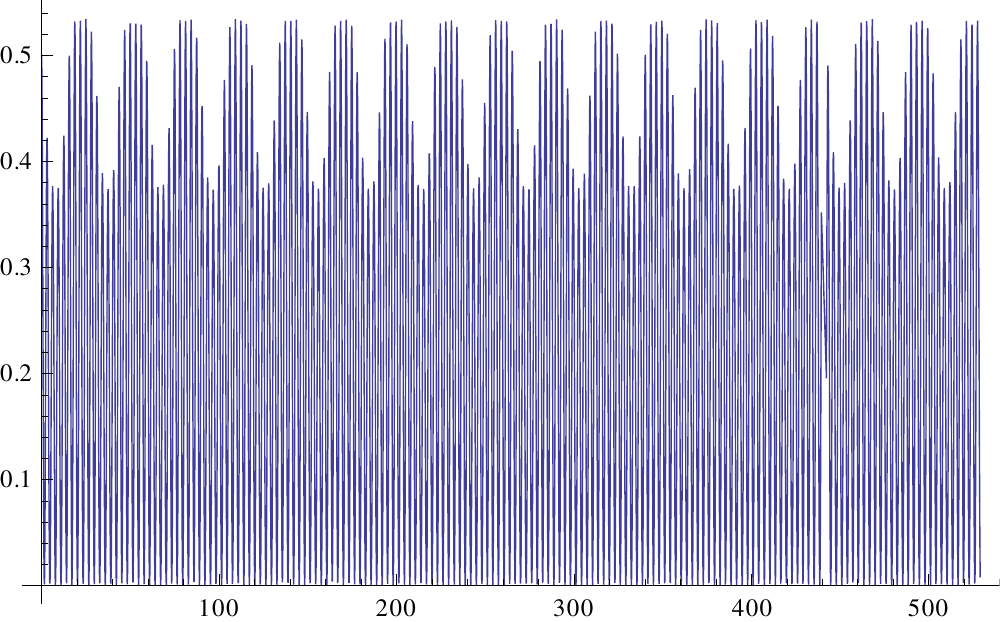}}
\put(25,-6){\includegraphics[scale=0.38]{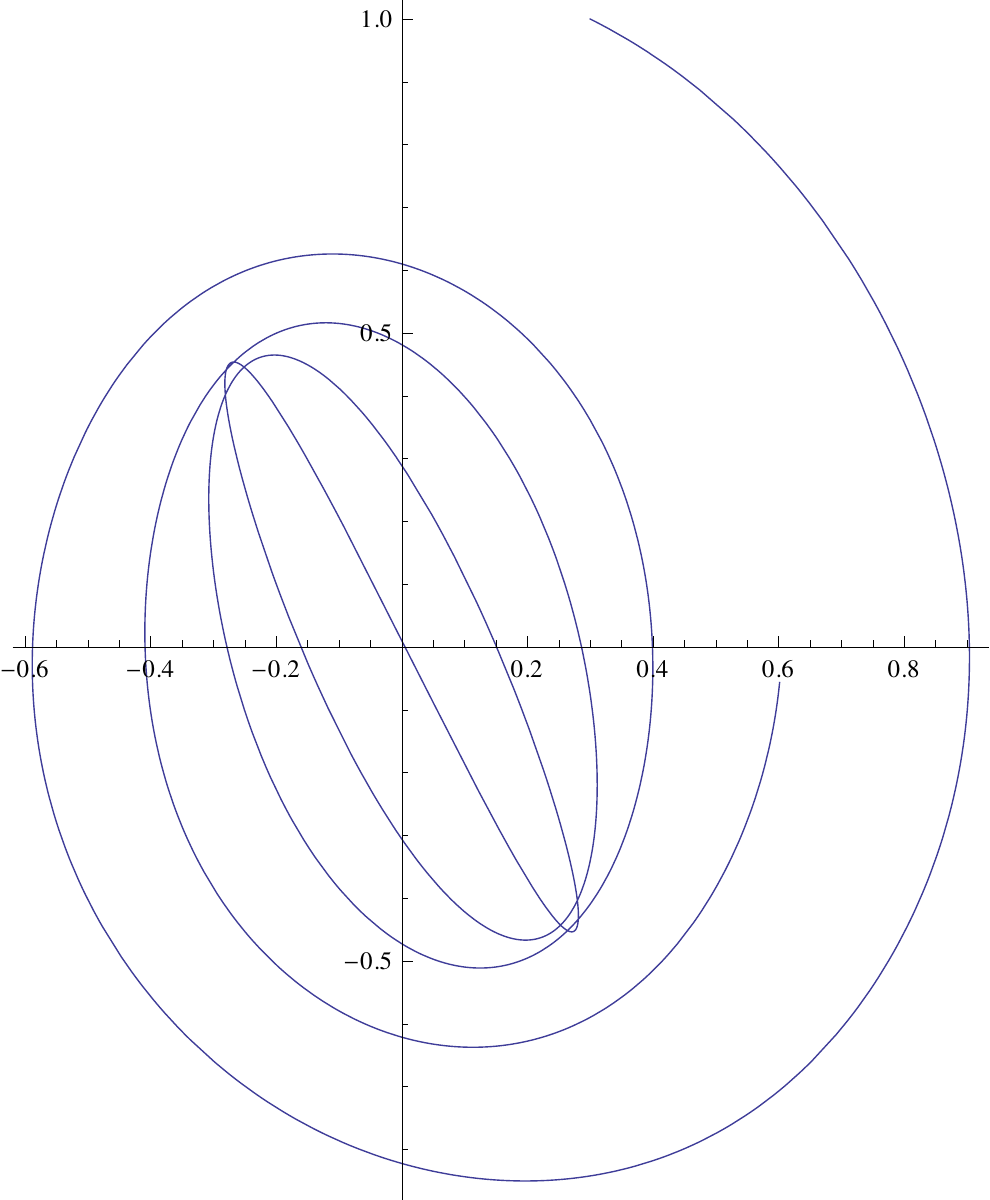}}
\put(90,0){\includegraphics[scale=0.4]{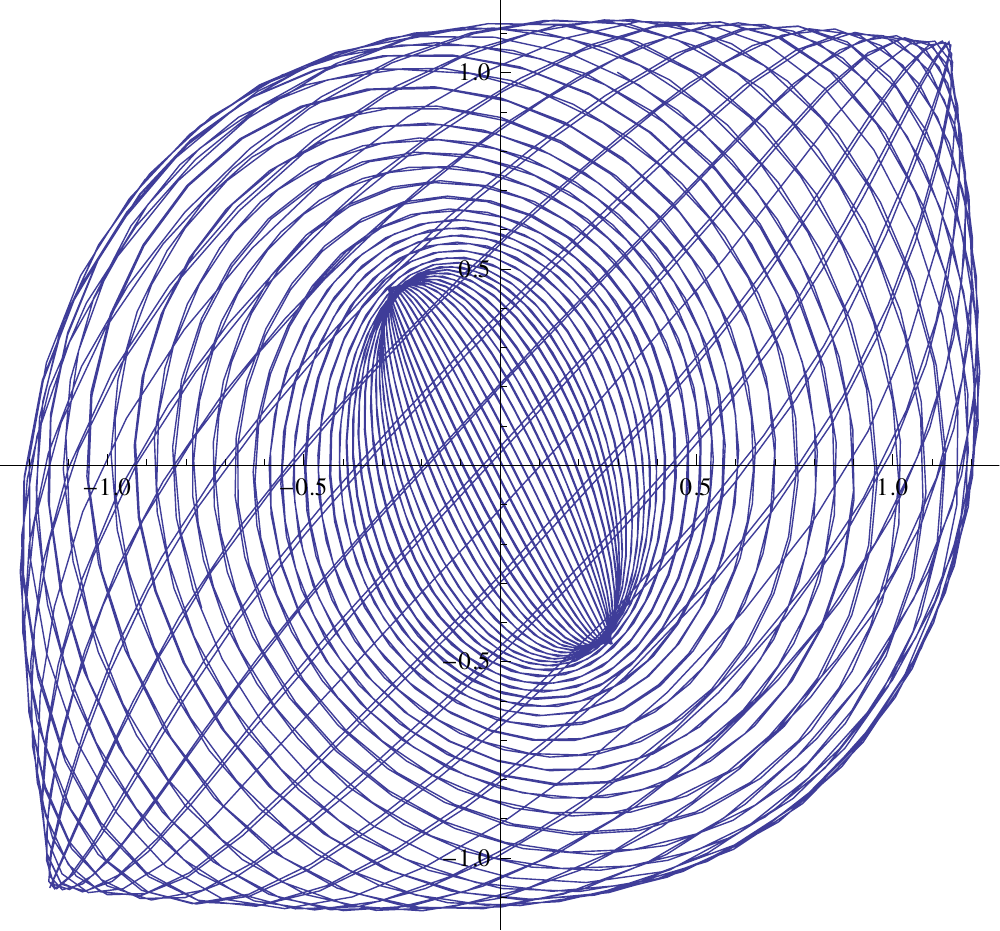}}
\put(150,0){\includegraphics[scale=0.4]{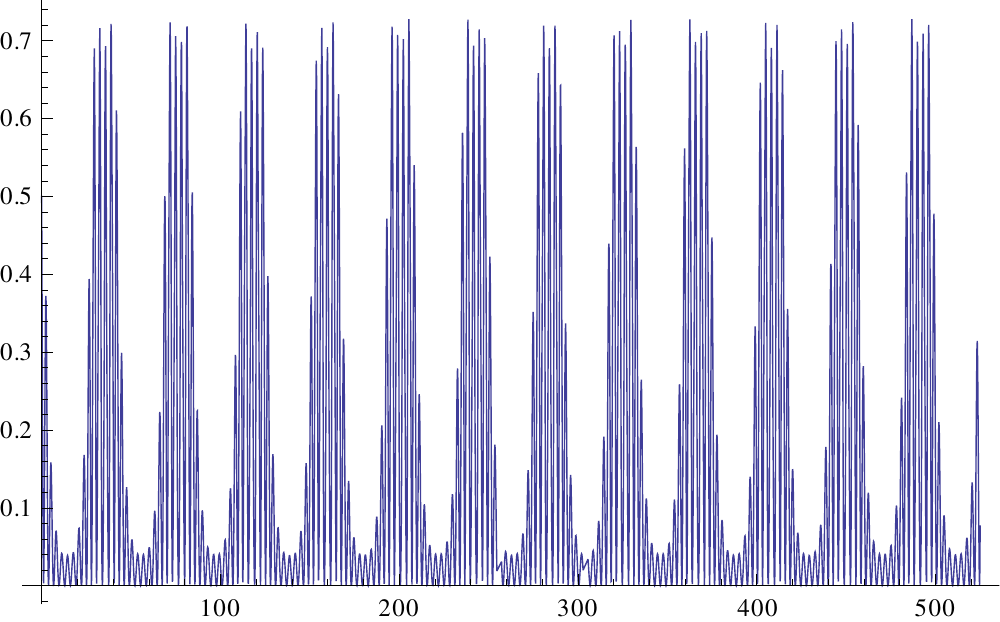}}

\put(76,81){$x'$}
\put(47,108){$y'$}

\put(141,79){$x'$}
\put(113,106){$y'$}

\put(73,22){$x'$}
\put(46,44){$y'$}

\put(141,20){$x'$}
\put(112,48){$y'$}

\put(147,89){$\frac{{\dot x}'^2}{2}$}
\put(201,53){$t$}

\put(147,33){$\frac{{\dot x}'^2}{2}$}
\put(201,-2){$t$}

\put(158,106){$^{\lambda=0.22}$}'
\put(157,101){$^{x'(0)=0,~y'(0)=1}$}
\put(157,96){$^{{\dot x}'(0)=1,~{\dot y}(0)=0}$}
\put(157,45){$^{\lambda=1}$}
\put(157,40){$^{x'(0)=0.3,~y'(0)=1}$}
\put(157,35){$^{{\dot x}'(0)=1,~{\dot y}'(0)=-0.5}$}

\end{picture}

\caption{\footnotesize Solutions to the equations of
motion (\ref{6.1a}), (\ref{6.1b}), in which the quartic
interaction $\frac{\lambda}{4} (x'+y')^4$
is replaced by $\frac{\lambda}{4}\text{sin}^4\, (x'+y')$. The system
is now stable for all positive values of $\lambda$.}
\end{figure}
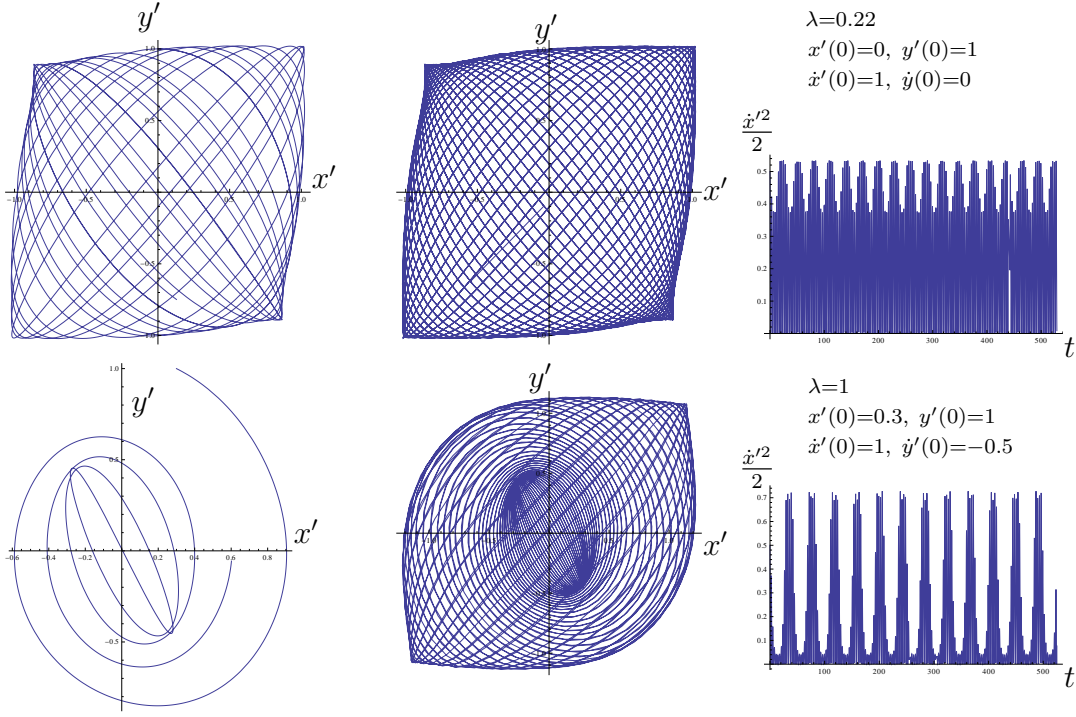

(ii) {\it Unequal masses}

The case when masses are different was also considered in Ref.\,\ci{PavsicPUstable}.
Instead of the Lagrangian (\ref{5.11}) we then have
\be
    L=\mbox{$\frac{1}{2}$}(m_1 {\dot x}'^2 - m_2 {\dot y}'^2) - \mbox{$\frac{1}{2}$}
    (\omega_1^2 x'^2 - \omega_2^2 y'^2) - \frac{\lambda}{4} (x'+y')^4
 \lbl{6.2}
\ee
Introducing
\be
     m= \mbox{$\frac{1}{2}$} (m_1 - m_2) , ~~~~ 
     M = \mbox{$\frac{1}{2}$} (m_1 + m_2),
\lbl{6.3}
\ee
and the variables $u$,$v$, defined in Eq.\,(\ref{5.4}), the Lagrangian (\ref{6.2})
becomes
\be
    L = \mbox{$\frac{1}{2}$} \left [ m ({\dot u}^2 + {\dot v}^2) + 2 M 
    {\dot u} {\dot v} + \rho_1 (u^2 + v^2) - 2 \mu_1 u v \right ]
    - \lambda u^4 ,
\lbl{6.4}
\ee
where $\mu_1$ and $\rho_1$ are defined in Eq.\,(\ref{5.6}).  This gives the
system of two coupled second order equation of motion
\be
    m {\ddot u} + M {\ddot v} - \rho_1 u + \mu_1 v + 4 \lambda u^3 = 0
\lbl{6.5}
\ee
\be
    m {\ddot v} + M {\ddot u} - \rho_1 v + \mu_1 u =0 .
\lbl{6.6}
\ee
which is equivalent to the fourth order equation
\be
u^{(4)} M (M^2-m^2) + 2 {\ddot u} M (\mu_1 M + \rho_1 m) 
+ u M (\mu_1^2 - \rho_1^2)
      + 4 M \rho_1\lambda u^3 - 4 M m \lambda \frac{\dd^2}{\dd t^2}
     \left ( u^3 \right ) = 0
\lbl{6.7}
\ee

By putting $M=1$, $m=0$, we verify that the latter equation becomes the
equation (\ref{5.15}) of the PU oscillator with the quartic self-Interaction term.
But if $M\neq 0$, $m\neq 0$, then Eq.\,(\ref{6.7}), because of the last term,
looks like a generalization of the equation of motion (\ref{5.10}) that was derived from the
Lagrangian (\ref{5.5}), equivalent to the Lagrangian (\ref{5.1}) with the signature
$(++)$ in the space of the variables $x'$,$y'$. This suggests that the system
with different masses is stable. Stability of such a system was confirmed\,\ci{PavsicPUstable}
by numerical solutions to the equation of motion
\be
    m_1 {\ddot x}' + \omega_1^2 x' + \lambda (x'+y')^3 = 0,
\lbl{6.8}
\ee
\be
     m_2 {\ddot y} + \omega_2^2 y' - \lambda (x'+y')^3 = 0,
\lbl{6.9}
\ee
derived from the Lagrangian (\ref{6.2}).

The results of many calculations, reported in Ref.\,\ci{PavsicPUstable} showed
stability for all values of $\lambda >0$ and initial velocities. Some examples are
shown in Fig.\,10. If masses are equal, $m_1=m_2 =1$, then the system of equations
(\ref{6.8}),(\ref{6.9}) with $\om_1=1$, $\om_2 = \sqrt{2}$ exhibits runaway
solution when $\lambda > 0.03$, whilst it is stable for lower values of
$\lambda$. But if masses are slightly different, then the system is stable.

Similar behaviour had been previously shown\,\ci{PavsicFirenze} for a similar
system, but with the coupling term $\frac{1}{4} (x^2-y^2)^2$ (see also Sec.\,4 of
this review).

Let us now show also analytically that the system of equations (\ref{6.8}),(\ref{6.9})
is stable for $m_1 <m_2$ and $\lambda \gg 0$. Assuming very high $\lambda$,
such that the contribution of the terms $\om_1^2 x'$ and $\om_2^2 y'$ are
negligible, the system (\ref{6.8}),(\ref{6.9}) becomes
\be
   {\ddot x}' + \frac{1}{m_1} \lambda (x'+y')^3 = 0,
\lbl{6.10}
\ee
\be
{\ddot y}' - \frac{1}{m_2} \lambda (x'+y')^3 = 0.
\lbl{6.11}
\ee
In terms of the new variables $\xi=x'+y'$ and $\eta=x'-y'$, the latter system reads
\be
   {\ddot \xi} + \left (\frac{1}{m_1} - \frac{1}{m_2} \right ) \lambda \xi^3=0,
\lbl{6.12}
\ee
\be
   {\ddot \eta} + \left (\frac{1}{m_1} + \frac{1}{m_2} \right ) \lambda \xi^3=0.
\lbl{6.13}
\ee
In the first equation only the variable $\xi$ occurs, and the quartic
potential is bounded from below if $m_1<m_2$ and $\lambda>0$. The solution
is stable. Consequently, also $\eta (t)$ in the second equation is stable.
This is true regardless of how high is $\lambda$.

If we do not neglect the terms $\om_1^2 x'$ and $\om_2^2 y'$, and rewrite
the equations (\ref{6.8}) and (\ref{6.9}) in terms of the variables
$\xi$, $\eta$, we obtain
\be
   {\ddot \xi} + \frac{1}{2} \left (\frac{\om_1^2}{m_1} + \frac{\om_2^2}{m_2} \right ) \xi
   + \frac{1}{2} \left (\frac{\om_1^2}{m_1} - \frac{\om_2^2}{m_2} \right ) \eta +
   \left (\frac{1}{m_1} - \frac{1}{m_2} \right ) \lambda \xi^3=0,
\lbl{6.14}
\ee
\be
   {\ddot \eta}+ \frac{1}{2} \left (\frac{\om_1^2}{m_1} - \frac{\om_2^2}{m_2} \right ) \xi
   + \frac{1}{2} \left (\frac{\om_1^2}{m_1} + \frac{\om_2^2}{m_2} \right ) \eta + 
   + \left (\frac{1}{m_1} + \frac{1}{m_2} \right ) \lambda \xi^3=0.
\lbl{6.15}
\ee
If
\be
   \frac{\om_1^2}{m_1} = \frac{\om_2^2}{m_2},
\lbl{6.16}
\ee
then the variable $\eta$ disappears from Eq.(\ref{6.14}), and what remains is the differential
equation for $\xi$ only, which has stable oscillatory solution if $\lambda>0$. Once we have
a solution for $\xi(t)$, we can plug it into Eq.\,(\ref{6.15}), which then becomes a harmonic
oscillator equation for the variable $\eta$ with a time dependent force term. The system
(\ref{6.14}),(\ref{6.15}), or equivalently (\ref{6.8}),(\ref{6.9}) is thus stable despite the fact
that it is derived from the Lagrangian (\ref{6.2}) in which the degrees of freedom $x$
and $y$ have positive and negative energy, respectively.

Numerical solutions shown in Fig.\,10 reveal that even if we relax the relation
(\ref{6.16}), the system is stable. Moreover, in Sec.\,4.2 it is shown that the system
in which the interacting potential is $\frac{\lambda}{4} (x^2-y^2)^2$ is also
stable, provided that the masses are different. Ilhan and Kovner\,\ci{Ilhan} studied
a system of the same form as in Eq.\,(\ref{4.1}), but with the interaction
potential $V_1 = \lambda_1 x^4-\lambda_2 y^4 + \mu x^2 y^2$, and they found
stability in many numerical runs.
\setlength{\unitlength}{.8mm}

\begin{figure}[h!]
\hs{3mm}
\begin{picture}(120,115)(25,0)

\put(25,61){\includegraphics[scale=0.4]{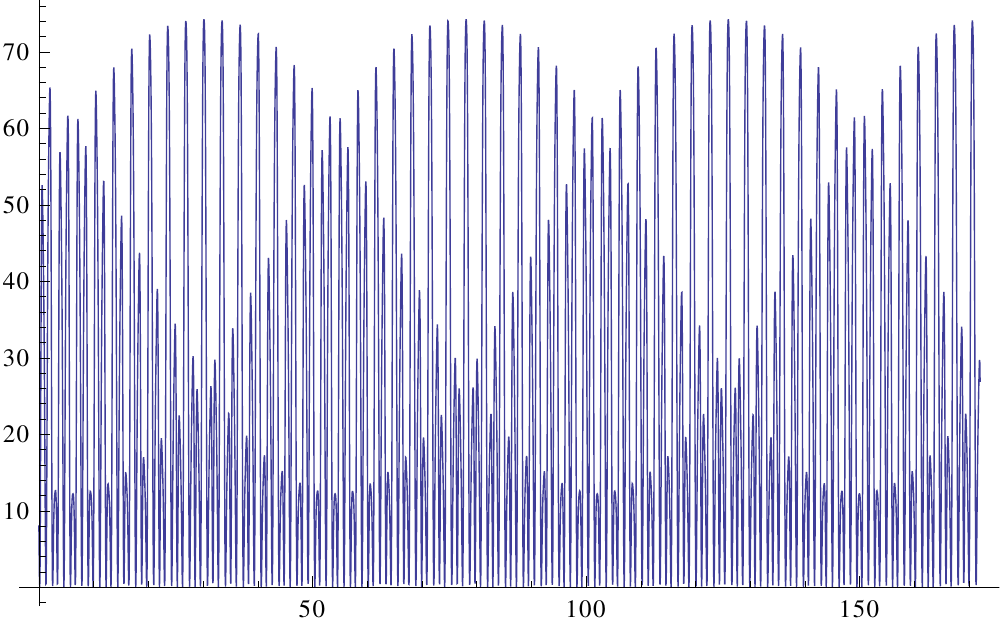}}
\put(90,61){\includegraphics[scale=0.4]{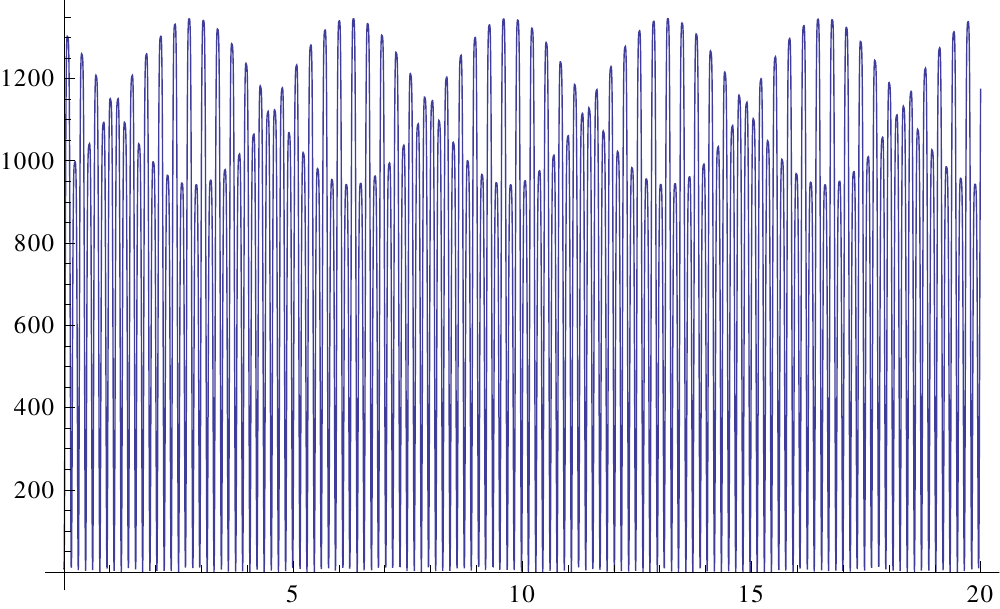}}
\put(150,61){\includegraphics[scale=0.4]{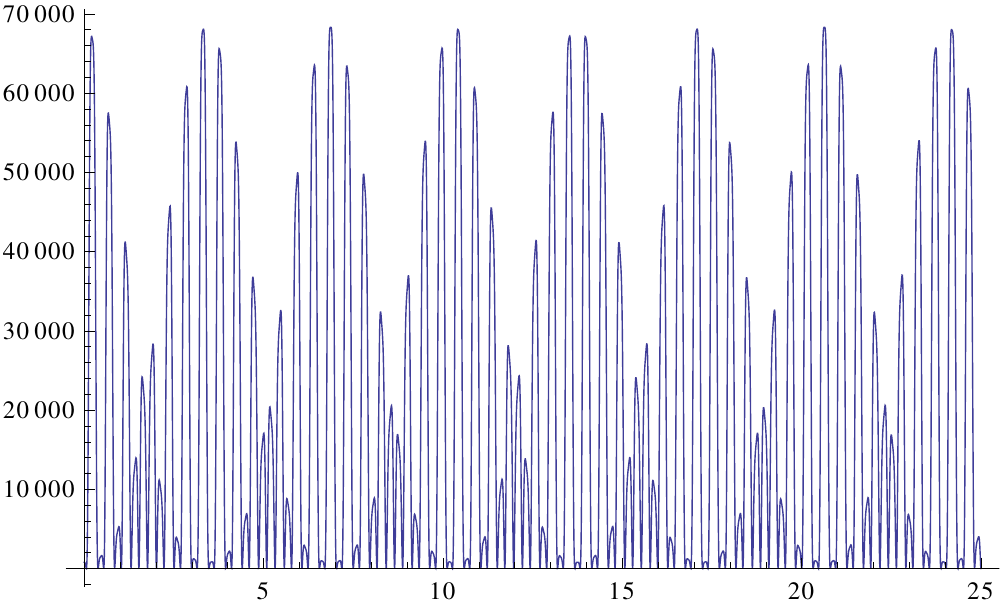}}
\put(25,0){\includegraphics[scale=0.4]{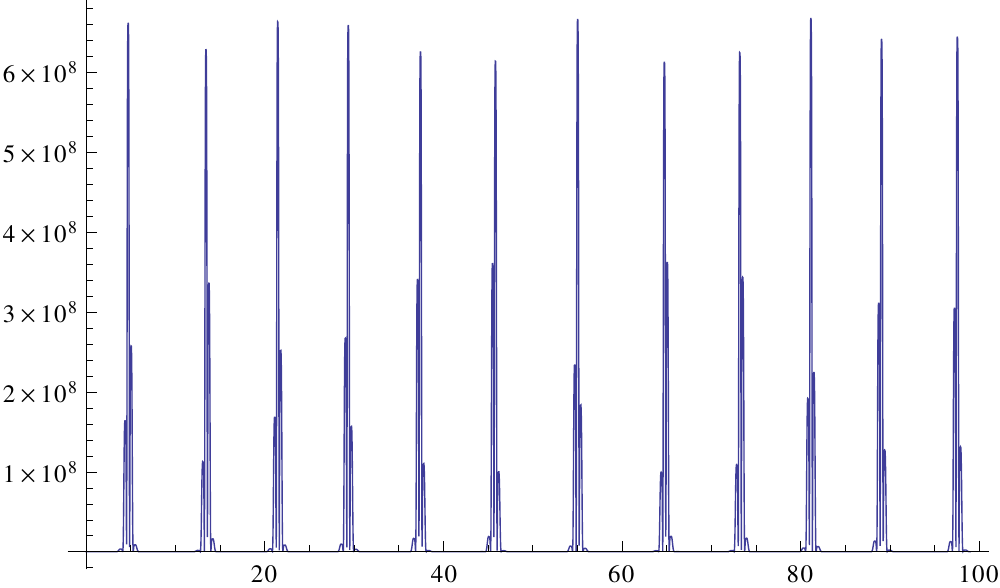}}
\put(90,0){\includegraphics[scale=0.4]{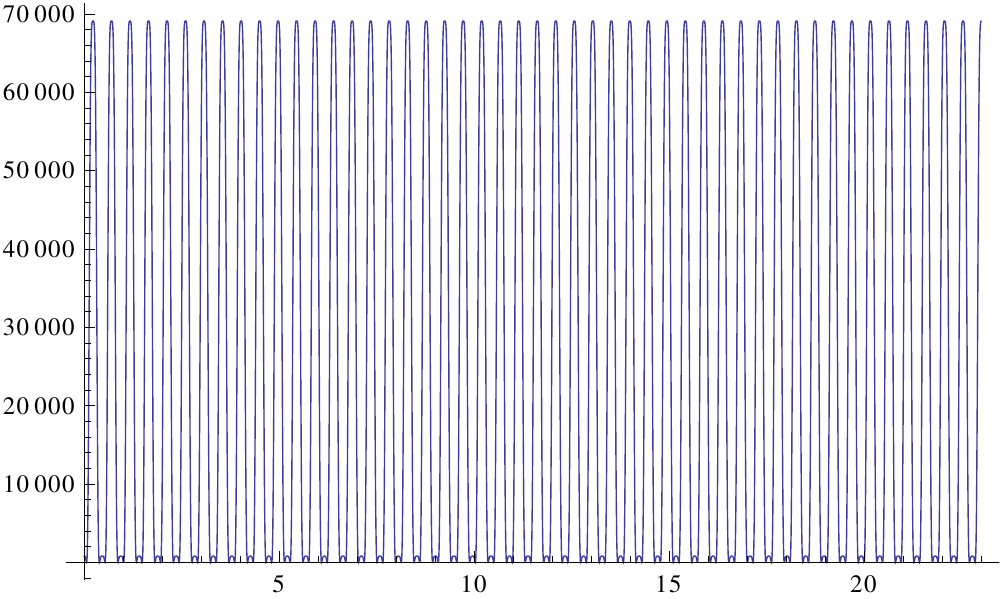}}
\put(150,0){\includegraphics[scale=0.4]{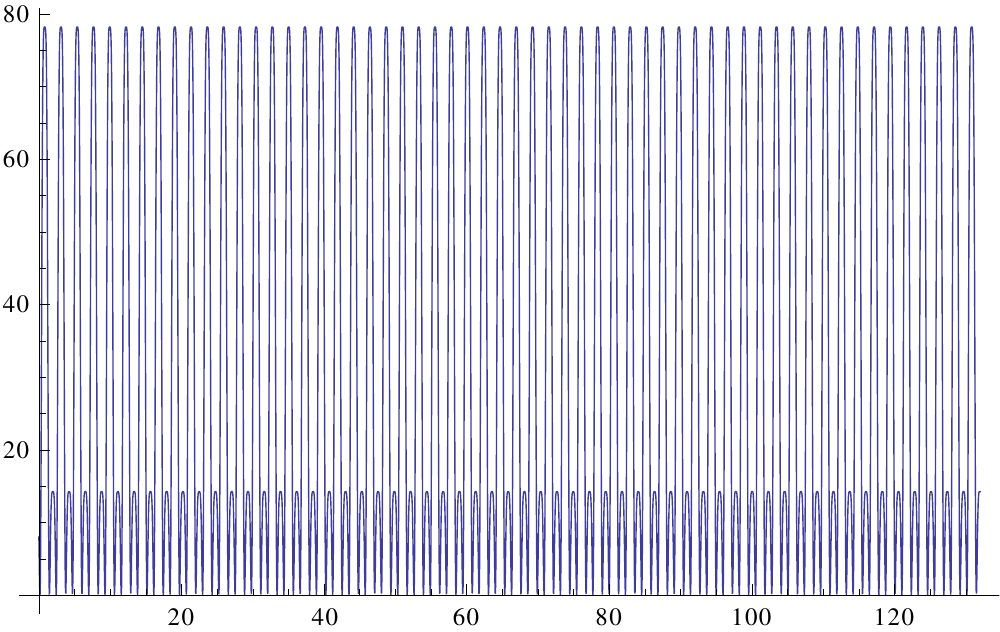}}

\put(21,92){$\frac{{\dot x}'^2}{2}$}
\put(76,59){$t$}

\put(87,92){$\frac{{\dot x}'^2}{2}$}
\put(141,59){$t$}

\put(145,92){$\frac{{\dot x}'^2}{2}$}
\put(201,59){$t$}

\put(21,30){$\frac{{\dot x}'^2}{2}$}
\put(76,-2){$t$}

\put(85,30){$\frac{{\dot x}'^2}{2}$}
\put(142,-2){$t$}

\put(145,30){$\frac{{\dot x}'^2}{2}$}
\put(202,-2){$t$}

\put(25,110){$^{\lambda=5,~m_1=0.7,~m_2=1.3}$}
\put(25,105){$^{x'(0)=0.3,~y'(0)=1}$}
\put(25,100){$^{{\dot x}'(0)=4,~{\dot y}'(0)=-0.5}$}
\put(30,95){$^{\omega_1=1,~\omega_2=\sqrt{1.5}}$}

\put(91,110){$^{\lambda=500,~m_1=0.7,~m_2=1.3}$}
\put(91,105){$^{x'(0)=0.3,~y'(0)=1}$}
\put(91,100){$^{{\dot x}'(0)=4,~{\dot y}'(0)=-0.5}$}
\put(96,95){$^{\omega_1=1,~\omega_2=\sqrt{1.5}}$}

\put(152,110){$^{\lambda=5,~m_1=0.7,~m_2=1.3}$}'
\put(152,105){$^{x'(0)=0.3,~y'(0)=1}$}
\put(152,100){$^{{\dot x}'(0)=40,~{\dot y}'(0)=55}$}
\put(157,95){$^{\omega_1=1,~\omega_2=\sqrt{1.5}}$}
\put(25,48){$^{\lambda=5,~m_1=0.99,~m_2=1.01}$}
\put(25,43){$^{x'(0)=0.3,~y'(0)=1}$}
\put(25,38){$^{{\dot x}'(0)=1,~{\dot y}'(0)=-0.5}$}
\put(30,33){$^{\omega_1=1,~\omega_2=\sqrt{1.5}}$}

\put(91,48){$^{\lambda=5,~m_1=0.7,~m_2=1.3}$}
\put(91,43){$^{x'(0)=0.3,~y'(0)=1}$}
\put(91,38){$^{{\dot x}'(0)=40,~{\dot y}'(0)=55}$}
\put(96,33){$^{\omega_1=\omega_2=0}$}

\put(152,48){$^{\lambda=5,~m_1=0.7,~m_2=1.3}$}
\put(152,43){$^{x'(0)=0.3,~y'(0)=1}$}
\put(152,38){$^{{\dot x}'(0)=4,~{\dot y}'(0)=-0.5}$}
\put(157,33){$^{\omega_1=\omega_2=0}$}

\end{picture}

\caption{\footnotesize Solutions of Eqs.\,(\ref{6.8})(\ref{6.9}) for
different values of the coupling constant $\lambda$ and different
initial conditions. We show here the kinetic energy ${\dot x}'^2/2$ as
function of time.}
\end{figure}
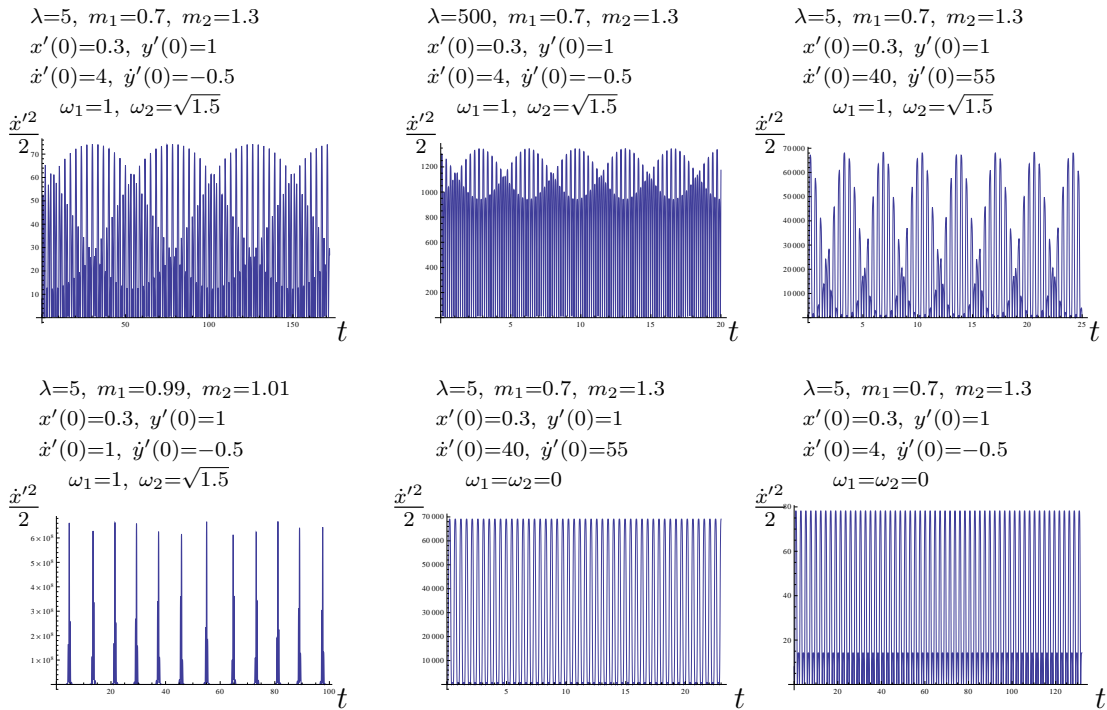

Though the system (\ref{4.1}) is equivalent to a self interacting PU oscillator
only if $V_1$ is a function of $x'+y'$ or $x'-y'$, and if $m_1=m_2$, the studies
revealing stability of the systems such as (\ref{4.1}) or (\ref{6.2}) are important,
because they show that the presence of negative energies does not
automatically destroy physical viability of such systems.

Returning now to the system of equations (\ref{6.12}),(\ref{6.13}) and
taking $m_1=m_2$, we find
\be
     {\ddot \xi}=0~,~~~~~~{\ddot \eta} + \frac{2}{m_1} \lambda \xi^3 =0,
\lbl{6.17}
\ee
with the general solution
\be  
    \xi=\xi_0 + c_1 t~,~~~~~~
 \eta = - \frac{2}{m_1} \frac{\lambda}{20} (\xi_0 + c_1 t)^5 + c_2 t    
\lbl{6.18}
\ee
which is a runaway trajectory. In the presence of the terms
$\omega_1^2 x$ and $\omega_2^2 y$, the above
runaway behavior is modulated by oscillations.

But if instead of $\frac{1}{4} (x'+y')^4$ we take the bounded
interaction potential $\frac{1}{4} {\rm sin}^4 (x'+y')$, then instead
of Eqs.\,(\ref{6.17}) we have
\be
     {\ddot \xi}=0~,~~~~~~{\ddot \eta} + \frac{2}{m_1} \lambda
     \,\text{sin}^3 \, \xi \, \text{cos}\, \xi =0,
\lbl{6.19}
\ee
whose general solution is
\bear  
   && \xi=\xi_0 + c_1 t~,~~~~~~~~~~{\dot \eta}= -\frac{2}{m_1} \frac{1}{4 c_1} \lambda
    \,\text{sin}^4 \,(\xi_0 + c_1 t)~,\nonumber\\
   && \eta = -\frac{2}{m_1} \frac{1}{128 c_1^2} \lambda
    \left [ 12 (\xi_0 + c_1 t) -8\, \text{sin}\,(2 (\xi_0 + c_1 t)
    +\text{sin}\, (4 (\xi_0 + c_1 t)  \right ]
\lbl{6.20}
\ear
This is a free particle trajectory modulated by finite oscillations.
The kinetic energy $\frac{m_1}{2}{\dot x}'^2 - \frac{m_1}{2}{\dot y}'^2=
\frac{m_1}{2}{\dot \xi}{\dot \eta}$ remains finite.

\section{Pais-Uhlenbeck oscillator in the presence of damping}

\subsection{Inclusion of a damping term}

An important factor that must be taken into account is the influence of
the environment, which manifests itself as dissipative forces, acting on a
system. In the case of an oscillator, the environment acts as a damping
force. Nesterenko\,\ci{Nesterenko} studied the behaviour of the
Pais-Uhlenbeck oscillator in the presence of a damping term that is 
linear in velocity. But Stephen showed\,\ci{Stephen}  that not only the
linear term ${\dot x}$, but also the term ${\dddot x}$ has to be present
in the PU oscilattor. Indeed, if we generalize the equation (\ref{3.2}) so
to included damping as well, we arrive at\,\ci{PavsicPUdamp}
 \be
    \left (\frac{\dd^2}{\dd t^2}+ 2 \beta \frac{\dd}{\dd t}+\omega_2^2 \right ) 
    \left (\frac{\dd^2}{\dd t^2}+2 \alpha \frac{\dd}{\dd t}+\omega_1^2 \right ) x 
    = 0,
\lbl{7.1} 
\ee  
which gives the following fourth order differential equation with two
damping terms
\be
 x^{(4)} + 2 (\alpha+\beta){\dddot x}
    +(\omega_1^2+\omega_2^2 +4 \alpha \beta) {\ddot x}
    +2 (\omega_1^2 \beta + \omega_2^2 \alpha) {\dot x} + \omega_1^2 \omega_2^2 x =0 , 
\lbl{7.2}
\ee
The general solution of the latter equation is\,\ci{PavsicPUdamp}
 \be
   x = {\rm e}^{- \alpha t} \left (C_1 {\rm e}^{t \sqrt{\alpha^2-\omega_1^2}}
       +C_2 {\rm e}^{- t \sqrt{\alpha^2-\omega_1^2}} \right )
     +{\rm e}^{- \beta t} \left (C_3 {\rm e}^{t \sqrt{\beta^2-\omega_2^2}}
       +C_4 {\rm e}^{- t \sqrt{\beta^2-\omega_2^2}} \right ) .
\lbl{7.3}
\ee
This is oscillatory function if $\alpha^2 < \omega_1^2$, $\beta^2 < \omega_2^2$.
If in addition $\alpha$ and $\beta$ are positive, then $x(t)$ has an exponentially
decaying envelop. Thus, under the above conditions, the solution (\ref{7.3})
is stable.

In particular, if $\alpha = - \beta$, then the solution (\ref{7.3}) has an
exponentially decreasing  and an exponentially increasing part:
\be
x = {\rm e}^{\beta t}\left (
   C_1 {\rm e}^{t \sqrt{\beta^2-\omega_1^2}} 
   +C_2 {\rm e}^{-t \sqrt{\beta^2-\omega_1^2}} \right )
    +{\rm e}^{- \beta t} \left ( C_3 {\rm e}^{t \sqrt{\beta^2-\omega_2^2}}
       +C_4 {\rm e}^{- t \sqrt{\beta^2-\omega_2^2}} \right ) .
\lbl{7.4}
\ee
Such a solution is thus unstable for every real $\beta$. If we plug 
$\alpha = - \beta$ into Eq.\,(\ref{7.2}) we obtain
\be
    x^{(4)} 
    +(\omega_1^2+\omega_2^2 -4 \beta^2) {\ddot x}
    +2 \beta (\omega_1^2 - \omega_2^2) {\dot x} + \omega_1^2 \omega_2^2\, x =0,
\lbl{7.5}
\ee    
which has no ${\dddot x}$ term. This can be rewritten into the form considered
by Nesterenko\,\ci{Nesterenko}:
\be
    x^{(4)} 
    +(\Omega_1^2+\Omega_2^2) {\ddot x}
    +2 \gamma {\dot x} + \Omega_1^2 \Omega_2^2 \, x =0,
\lbl{7.6}
\ee    
where $\gam = \beta (\omega_1^2 - \omega_2^2)$, and
\be
    \Omega_1^2+\Omega_2^2 = \omega_1^2+\omega_2^2 - 4 \beta^2,
\lbl{7.7}
\ee
\be
     \Omega_1^2 \Omega_2^2 = \omega_1^2 \omega_2^2 .
\lbl{7.8}
\ee
The latter system of equations has the solution
\be
    \Omega_{1,2}^2 = \frac{1}{2} \left [ \omega_1^2 + \omega_2^2 - 4 \beta^2
      \pm \sqrt{(\omega_1^2 + \omega_2^2 - 4 \beta^2)^2- 4 \omega_1^2 \omega_2^2}
      \right ] .
\lbl{7.9}
\ee

Nesterenko studied an approximate solution to Eq.\,(\ref{7.6}) for weak
dumping, by using the perturbation theory. His results is in agreement with
the exact solution (\ref{7.4}) considered in Ref.\,\ci{PavsicPUdamp}, which in
turn is a special case (for $\alpha = - \beta$) of the solution (\ref{7.3}) of the
oscillator with the ${\dot x}$ and ${\dddot x}$ damping terms.

\subsection{Presence of an arbitrary external force}

Nesterenko\,\ci{Nesterenko} considered also the Pais-Uhlenbeck oscillator
which experiences an arbitrary external force $f(t)$. He expressed the
solution $x(t)$ in terms of the propagator $G(t-t')$, or equivalently, in terms
of its Fourier transform ${\tilde G}(\om)$. He found that ${\tl G}(\om)$
contains a positive and a negative term, and thus the displacement
$x(t)$ consists of two contributions. Nesterenko concluded that one of
these contributions to $x(t)$ was unphysical.

In Ref.\,\ci{PavsicPUdamp} and arbitrary time dependent force $f(t)$ was
added to the equation of motion (\ref{7.1}) for the PU oscillator with
damping:
 \be
    \left (\frac{\dd^2}{\dd t^2}+ 2 \beta \frac{\dd}{\dd t}+\omega_2^2 \right ) 
    \left (\frac{\dd^2}{\dd t^2}+2 \alpha \frac{\dd}{\dd t}+\omega_1^2 \right ) x 
    = f(t).
\lbl{7.10}
\ee
Two cases, (i) without damping, $\alpha=\beta=0$, and (ii) with damping,
$\alpha\neq 0$, $\beta\neq0$, were considered in some more details\,\ci{PavsicPUdamp}.

The general solution of Eq.\,(\ref{7.10}) is
\bear
     &&x(t)= x_0 (t) +\int_{-\infty}^{\infty} G(t-t') f(t') \dd t' \nonumber\\
     &&~~~~~=x_0(t) + \int_{-\infty}^{\infty} {\tl G}(\om) {\tl f}(\om) \dd \om,
\lbl{7.11}
\ear
where $x_0(t)$ is a general solution of the homogeneous equation (\ref{7.1}),
i.e., Eq.\,(\ref{7.3}), and where
\be
      G(t) = \frac{1}{\sqrt{2 \pi}} \int_{-\infty}^{\infty} 
      {\rm e}^{i \omega t} {\tl G} (\omega) \dd \omega ,
\lbl{7.13}
\ee
\be
      f(t) = \frac{1}{\sqrt{2 \pi}} \int_{-\infty}^{\infty} 
      {\rm e}^{i \omega t} {\tl f} (\omega) \dd \omega ,
\lbl{7.14}
\ee

In {\it Case (i)}, the Fourier transformed Green function, derived from
Eq.\,(\ref{7.10}) for $\alpha=\beta=0$, is
\be
   {\tl G} (\omega) = \frac{1}{\sqrt{2 \pi}(\omega_1^2 - \omega_2^2)}
   \left (\frac{1}{\omega^2 - \omega_1^2}-\frac{1}{\omega^2 - \omega_2^2}
   \right ) ,
\lbl{7.15}
\ee
which for $t>0$ gives
\be
   G(t) = \frac{1}{2(\omega_1^2 - \omega_2^2)}
   \left ( \frac{{\rm sin}\,\omega_2 t}{\omega_2} -
          \frac{{\rm sin}\,\omega_1 t}{\omega_1} \right ) .
\lbl{7.15a}
\ee

As an example, the following force was considered\,\ci{PavsicPUdamp}:
\be
   f(t) = a \,{\rm cos}\,\omega_1 t + b\, {\rm cos}\,\omega_2 t ,
\lbl{7.16}
\ee
which according to Eq.\,(\ref{7.14}) gives
\be
{\tl f}(\omega)=\frac{a \sqrt{2 \pi}}{2}[\delta(\omega-\omega_1)
+\delta(\omega+\omega_1)] + \frac{b \sqrt{2 \pi}}{2}[
\delta(\omega - \omega_2)+\delta(\omega + \omega_2)].
\lbl{7.17}
\ee
Using (\ref{7.15a}) and (\ref{7.16}), the solution (\ref{7.11}) reads explicitly
  $$x(t) = x_0 (t) - \frac{1}{2 \omega_1 \omega_2 (\omega_1^2 - \omega_2^2)^2}
   \Bigl[ (\omega_1^2 - \omega_2^2) (a \omega_2 t \,{\rm sin}\,\omega_1 t
   - b \omega_1 t \,{\rm sin}\,\omega_2 t)  \hs{2cm}$$
\be   
  \hs{2.5cm} ~+ 2 (a-b) \omega_1 \omega_2 ({\rm cos}\,\omega_1 t - {\rm cos}\,\omega_2 t)
   \Bigr] ,
\lbl{7.18}
\ee
where $x_0 (t)$ is given by Eq.\,(\ref{7.3}) in which we set $\alpha=0$, $\beta=0$.
This solution was also obtained directly from Eq.\,(\ref{7.10}) by using  the command
DSolve in Mathematica.

In {\it Case (ii)} we have
\be
   {\tl G}(\om) = \frac{1}{\sqrt{2 \pi}} \frac{1}{(\om^2-\om_2^2 - 2 i \beta)(\om^2-\om_1^2-2 i \alpha)}
\lbl{7.19}
\ee
Then the relation (\ref{7.11}) is
\bear
  &&x(t) =x_0 (t) + \int  {\tl G} (\om) {\tl f}(\om) \dd \om \nonumber\\
  &&\hs{8mm}=x_0 (t) + \frac{a}{4 i \alpha \om_1} \left ( \frac{{\rm e}^{i \om_1 t}}{-\om_1^2 
  + \om_2^2 +2 i \beta \om_1}
-  \frac{{\rm e}^{-i \om_1 t}}{-\om_1^2 + \om_2^2 -2 i \beta \om_1} \right )\nonumber \\
  &&\hs{22mm}+\frac{b}{4 i \beta \om_2} \left ( \frac{{\rm e}^{i \om_2 t}}{-\om_2^2 + \om_1^2 +2 i \alpha \om_2}
-  \frac{{\rm e}^{-i \om_1 t}}{-\om_2^2 + \om_1^2 -2 i \alpha \om_2} \right )\nonumber \\
  &&\hs{8mm}=   x_0 (t) + \,\frac{-2 \beta \omega_1 a{\rm cos}\, \omega_1 t -
    a (\omega_1^2-\omega_2^2) {\rm sin}\,\omega_1 t}
    {2 \alpha \omega_1 [4 \beta^2 \omega_1^2 +(\omega_1^2-\omega_2^2)^2]}\nonumber\\
    &&\hs{22mm}+ \frac{-2\alpha \omega_2 b{\rm cos}\, \omega_2 t +
    b (\omega_1^2-\omega_2^2) {\rm sin}\,\omega_2 t}
    {2 \beta \omega_2 [4 \alpha^2 \omega_2^2 +(\omega_1^2-\omega_2^2)^2]},
\lbl{7.20}
\ear
where $x_0 (t)$ is given by Eq.\,(\ref{7.3}) for $\alpha>0$, $\beta>0$. The same solution (\ref{7.20})
can be obtained by applying the Mathematica command DSolve to the 4th order differential
equation (\ref{7.10}) with $f(t)$ given in Eq.\, (\ref{7.16}), which confirmed the correctness of the above
result.

We see from Eq.\,(\ref{7.18}) that the amplitude in the solution $x(t)$ for $\alpha=\beta=0$
increases linearly with time $t$, and thus such a system is unstable. This is not the case in the
presence of damping, i.e., when $\alpha>0$, $\beta>0$. Then the solution is (\ref{7.20}) which
exhibits decent oscillatory behaviour. Here the behaviour of the displacement $x(t)$ is very
similar as in the case of a harmonic oscillator in the presence of a force that oscillates with
the resonant frequency. In the absence of damping, the amplitude of $x(t)$ grows linearly
with time, whereas in the presence of damping it remains confined.

In Eq.\,(\ref{7.16}) we considered the time dependent force $f(t)$ whose spectral
density is sharply localized around $\om_1^2$ and $\om_2^2$ according to Eq.\,(\ref{7.17}).
In Ref.\,\ci{PavsicPUdamp} also a more general case
\be
    {\tl f} (\omega) = \frac{a \sqrt{2 \pi}}{2} \sqrt{\frac{c}{\pi}} 
    \left ( {\rm e}^{-c(\omega - \omega_1)^2}
    +{\rm e}^{-c(\omega + \omega_1)^2} \right )
  + \frac{b \sqrt{2 \pi}}{2} \sqrt{\frac{c}{\pi}} 
  \left ( {\rm e}^{-c(\omega - \omega_2)^2} 
  + {\rm e}^{-c(\omega + \omega_2)^2} \right ),
\lbl{7.21}
\ee
was considered. Then the time dependent force has exponentially decreasing
envelop:
\be
     f(t) = {\rm e}^{-\frac{t^2}{4 c}}
     (a \,{\rm cos}\,\omega_1 t + b\, {\rm cos}\,\omega_2 t ) .
\lbl{7.22}
\ee
In such a case the system is stable even in the absence of damping, i.e., when
$\alpha=\beta=0$. In Ref.\,\ci{PavsicPUdamp}, numerical solutions of
Eq.\,(\ref{7.10}) were found by using the Mathematic command NDSolve, and
they were stable for all finite positive values of the width parameter $c$ and initial velocities.
Some examples are shown in Fig.\,11.
\setlength{\unitlength}{0.8mm}

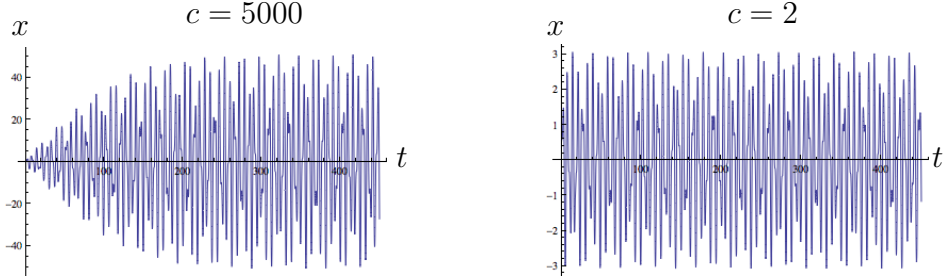
\begin{figure}[ht]
\hs{3mm} \begin{picture}(100,44)(-15,0)

\put(0,0){\includegraphics[scale=0.4]{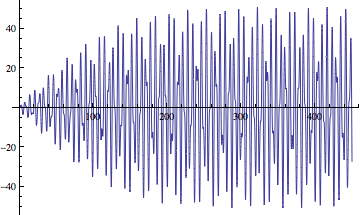}}
\put(90,0){\includegraphics[scale=0.4]{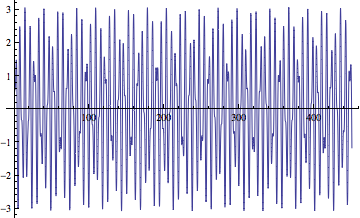}}

\put(65,18){$t$}
\put(154,18){$t$}
\put(1,40){$x$}
\put(90,40){$x$}
\put(30,42){$c=5000$}
\put(120,42){$c=2$}

\end{picture}

\caption{\footnotesize Solution of the undamped Pais-Uhlenbeck oscillator
($\alpha=\beta=0$) in the presence
of an external force with the spectral density localized around $\omega_1^2=1$
and $\omega_2^2=3$ according to (\ref{7.21}) for two different values of the width
parameter $c$. We took the constants $a=b=1$,
and the initial conditions $x(0)=0.5$, ${\dot x}(0)=0.4$, ${\ddot x}(0)=-1.2$, ${\dddot x}(0)=1$.}

\end{figure}

Another possibility is to consider the external force of the form
\be
 f(t) = a \, {\rm cos} \, \om_1' t + b\,  {\rm cos}\, \om'_2 t ,
\lbl{7.23}
\ee
where the frequencies $\om'_1$, $\om'_2$ are different from the resonant frequencies
$\om_1$, $\om_2$. Then in the case $\alpha=\beta=0$, the exact solution is
\be
  x(t) = x_0(t) -\frac{{\rm cos} \,\om'_1 t}{(\om_1^2-{\om'}_1'^2)({\om'}_1^2-\om_2^2)}
  -\frac{{\rm cos} \,\om'_1 t}{(\om_1^2-{\om'}_2'^2)({\om'}_2^2-\om_2^2)}
  \lbl{7.24}
\ee
In the case $\alpha\neq 0$, $\beta \neq 0$, we obtain the following exact solution:
\bear
   &&x(t) = x_0(t) 
    - \left [ \left (\om_1^2 \om_2^2-(4 \alpha \beta +\om_1^2 +\om_2^2) {\om'}_1^2 +{\om'}_1^4 \right )
   \left ( (\om_1^2 - {\om'}_2^2 )^2 + 4 \alpha^2 {\om'}_2^2 \right ) \right . \nonumber \\
  &&\hs{6cm} \times \left ( (\om_2^2 - {\om'}_2^2 )^2 + 4 \beta^2 {\om'}_2^2 \right ) {\rm cos}\, \om'_1 t \nonumber \\
  ~~~&&\hs{2cm} - \left ( \om_1^2 \om_2^2-(4 \alpha \beta +\om_1^2 +\om_2^2) {\om'}_2^2 +{\om'}_2^4 \right )
   \left ( (\om_1^2 - {\om'}_1^2 )^2 + 4 \alpha^2 {\om'}_1^2 \right ) \nonumber \\
  && \hs{6cm}\times \left ( (\om_2^2 - {\om'}_1^2 )^2 + 4 \beta^2 {\om'}_1^2 \right ) {\rm cos}\, \om'_2 t 
  \nonumber \\
 &&\hs{2cm} - 2 \om'_1 \left (\beta \om_1^2+\alpha \om_2^2 - (\alpha+\beta){\om'}_1^2 \right )
  \left ( (\om_1^2 - {\om'}_2^2)^2 + 4 \beta^2 {\om'}_2^2 \right ) \nonumber \\
  &&\hs{6cm} \times \left ( (\om_2^2-{\om'}_2^2)^2+4 \beta^2 {\om'}_2^2 \right ) {\rm sin}\,\om'_1  t \nonumber\\
 && \hs{2cm}-2 \om'_2 \left (\beta \om_1^2+\alpha \om_2^2 - (\alpha+\beta){\om'}_2^2 \right )
  \left ( (\om_1^2 - {\om'}_2^2)^2 + 4 \alpha^2 {\om'}_2^2 \right ) \nonumber \\
  &&\hs{6cm}\times  \left . \left ( (\om_2^2-{\om'}_1^2)^2+4 \beta^2 {\om'}_1^2 \right ) {\rm sin}\, \om'_2 t 
  \right ] \nonumber\\
  &&\times \left [ \left (\om_1^4 - 2 \om_1^2 {\om'}_1^2 + {\om'}_1^2 (4 \alpha^2+{\om'}_1^2) \right )
  \left ( \om_2^4 - 2 \om_2^2 {\om'}_1^2 +{\om'}_1^2 (4 \beta^2+{\om'}_1^2) \right ) \right . \nonumber\\
  &&\left . ~\times \left (\om_1^4 - 2 \om_1^2 {\om'}_2^2 + {\om'}_2^2 (4 \alpha^2+{\om'}_2^2) \right )
  \left ( \om_2^4 - 2 \om_2^2 {\om'}_2^2 +{\om'}_2^2 (4 \beta^2+{\om'}_2^2) \right ) \right ]^{-1}
\lbl{7.25}
\ear
In both cases we used Mathematica's DSolve, and verified that the solution so obtained indeed
satisfied the 4th order differential equation (\ref{7.10}) that we started from.

Nesterenko\,\ci{Nesterenko} remarked that due to the expression (\ref{7.15}) for the propagator
the forces with spectral densities localized around $\om_1^2$ and $\om_2^2$ give rise to
displacement $x$ of opposite signs. According to Nesterenko this implies that one of
these displacements is unphysical. But our calculations reveal that nothing unphysical happens
with the classical displacement $x(t)$.  The displacement rises into infinity in the physically
unrealistic case of the $\delta$-function localization of ${\tl f}(\om)$, i.e., when the width
parameter $c$ in Eq.\,(\ref{7.21}) and (\ref{7.22}) is infinite, i.e., when $f(t)$ is a sum
(up to a phase) of pure ${\rm cos} \,\om_1 t$ and ${\rm cos} \, \om_2 t$. In practice this is
impossible, because a realistic time dependent force cannot have {\it exactly} the
resonant frequencies  $\om_1$ and $\om_2$, the frequencies are at least slightly
different, or the spectral density is spread around them.
Moreover, there is always some damping. In either case, the solution is stable. The contributions of both terms in the Green's function ${\tl G} (\om)$ together give rise
to a physical displacement.

\section{On the stability of higher derivative field theories}

A higher derivative field theory of the form (\ref{2.1}) is an infinite set of Pais-Uhlenbeck
oscillators. If instead of the interaction potential $\lambda \phi^4$ we take a potential
$V_1(\phi)$ that is bounded from below and form above, then just as in the case
of the PU oscillator also such a field theoretic system is stable. However, now that we
have infinite dimensional system, the number of available final states, and hence the phase
space, can be infinite. In the studies of physical processes, one usually sums
over final states. With both positive and negative energies, the phase space can
be infinite. The formulas for transition and decay rates thus besides the transition
amplitude squared contain the integration over the phase space. A common
conclusion is that because in the case of higher derivative theory the phase space
is infinite, vacuum would instantaneously decay into positive and negative energy
particles, which means that such theories are not physically viable.

Such a conclusion has been questioned in Ref.\,\ci{PavsicFirenze}. When studying a
physical system, for instance an excited nucleus, we usually do not measure
the momenta of the decay particles, therefore in the transition probability formula
\be
   \text{Transition probability} = \int |S_{fi}|^2 \dd^3 \bp_1 \dd^3 \bp_2 ...
\lbl{8.0}
\ee
we integrate over them. If we measure the final state particle's momenta,
the integration goes over a narrower portion of the phase space, which
reduces the transition probability.

As an example let us consider the scalar fields described by the action
\be
   I = \frac{1}{2}\int {\dd^4 x [g^{\mu \nu } \partial _\mu  \varphi ^a 
   \partial _\nu  \varphi ^b \gam_{ab}  + V(\varphi )]} , 
\lbl{8.1}
\ee
If the metric $\gam_{ab}$ in the space of the fields $\varphi^a$ is indefinite with signature
$(r,s)$, then the latter action is a generalization of the action (\ref{2.3}) that comes from the higher
derivative action (\ref{2.1}). Let  the potential be of the form
\be
  V(\varphi)= m^2 \varphi^a \varphi^b \gam_{ab}
  + \varphi^a \varphi^b \varphi^c \varphi^d \lambda_{abcd},
\lbl{8.2}
\ee
or similar.

After performing the standard quantization procedure, the system is described by a state
vector $|\Psi \rangle$ expanded in terms of the Fock space basis vectors, $|P \rangle \equiv |p_1 p_2 ... p_n \rangle$ that are eigenvectors of the free field Hamiltonian $H_0$
(without the interaction quartic term $\varphi^a \varphi^b \varphi^c \varphi^d \lambda_{abcd}$):
\be
|\Psi \rangle  = \sum\limits_{} {|P\rangle \,\langle P|\Psi \rangle } ,
\lbl{8.3}
\ee
The system evolves according to the Schr\"odinger equation
\be
  i \frac{\p |\Psi \rangle}{\p t}=H |\Psi \rangle ,
\lbl{8.4}
\ee
with the formal solution
\be
  |\Psi (t)\rangle  = e^{ - iH(t - t_0 )} |\Psi (t_0 )\rangle .
\lbl{8.5}
\ee
Because the signature of the field space metric $\gam_{ab}$ is $(r,s)$, a Fock space state vector
contains particles with positive and negative energies. A consequence is that if the initial state
is the vacuum $|\psi (t_0) \rangle =|0 \rangle$, it can evolve into a superposition of many particle states, because the transitions
\be
 \langle P|e^{ - iH(t - t_0 )} |0\rangle = \langle P|\Psi (t)\rangle 
\lbl{8.6}
\ee
can satisfy the energy and momentum conservation.

The vacuum thus decays into a superposition of many particle states:
\be
 |\Psi (t)\rangle  = \sum\limits_{n = 0}^\infty  
 {|p_1 p_2 ...p_n \rangle \,\langle p_1 p_2 ...p_n |\Psi (t)\rangle } ,
\lbl{8.7}
\ee
where $\langle p_1 p_2 ...p_n |\Psi (t)\rangle$ is the probability amplitude of observing
at time $t$ the many particle state $|p_1 p_2 ...p_n \rangle$. The probabilities
that the vacuum decays into any of the states $|p_1 \rangle$,
$|p_1 p_2 \rangle$, $|p_1 p_2 ...p_n \rangle$, ...,
are not very different from each other, and, after a proper normalization, they sum to
$1-|\langle 0|\Psi \rangle |^2$. We thus have
\be
  |\langle 0|\Psi \rangle |^2 + \sum\limits_{p_1 } {|\langle p_1 |\Psi \rangle |^2  + } 
\sum\limits_{p_1 ,p_2 } {|\langle p_1 ,p_2 |\Psi \rangle |^2  + } 
\sum\limits_{p_1 ,p_2 ,...,p_n } {|\langle p_1 ,p_2 ,...,p_n |\Psi \rangle |^2
  + } \,...\,\, = \,1 .
\lbl{8.8}
\ee
This is so, because at any time $t$ the system must be in one of the states $|0 \rangle$,
$|p_1 \rangle $, $|p_1,p_2 \rangle$, ...., $|p_1,p_2,...,p_n \rangle $. Therefore the total
probability of finding the system in any of those states is $1$. However, the probability that
vacuum decays into 2,4,6,8, or any finite number of particles is infinitely small in comparison
with the probability that it decays into infinite numbers of particles, because such configurations
occupy the vast  majority of the phase space. Therefore, for an {\it outside observer},
who does not measure momenta of the particles, the vacuum $| 0 \rangle$ instantly decays
into infinitely many particles. The usual reasoning in the literature then goes along the lines
that because of such vacuum instability the theories involving ultrahyperbolic space, and higher
derivative theories in particular, are not physically viable.

However, every reasoning is based on certain assumptions, often implicit or tacit.
The reasoning against higher derivative field theories based on the instantaneous
vacuum decay due to the infinite phase space tacitly assumes the existence of
an observer, outside the system described by the higher derivative field theory,
who does not measure the number of particles and their momenta. But in reality,
such an outside observer cannot exist. Within our universe there can be no observer to whom the higher
derivative theory does not apply, if such a theory is the theory describing our universe.
Every conceivable observer is thus  a part of our universe. If so, he is then coupled to the particles in the
universe, and thus he at least implicitly measures their number and momenta. For such
an observer, there is no instantaneous vacuum decay for the reason discussed
in Ref.\,\ci{PavsicFirenze}, quoted below:

\begin{quotation}
Let us consider a generalization of the field action (\ref{8.1}).
We can rewrite it in a more compact notation:
\be
I = \frac{1}{2}\partial _\mu  \varphi ^{a(x)} \partial _\nu  \varphi ^{b(x')} 
\gamma _{\,\,\,\,\,\,\,\,a(x)b(x')}^{\mu \nu }  - U[\varphi ]
\lbl{q5.33}
\ee
Here $\varphi^{a(x)} \equiv \varphi^a (x)$, where ${(x)}$ is the continuous
index, denoting components of an infinite dimensional vector. In addition,
for every $(x)$, the components are also denoted by a discrete index $a$.
Altogether, vector components are denoted by the double index $a(x)$.
Alternative notation, often used in the literature, is
$\varphi^a (x) \equiv \varphi^{ax}$ or $\varphi^a (x) \equiv \varphi^{(ax)}$.

The action (\ref{q5.33}) may be obtained from a higher dimensional action
\be
I_\phi   = \frac{1}{2}\partial _\mu  \phi ^{A(x)} \partial _\nu  
\phi ^{B(x')} \,G_{\,\,\,\,\,A(x)B(x')}^{\mu \nu } ,
\lbl{q5.34}
\ee
where $A=(a,{\bar A})$, and $\phi^{A(x)} =(\phi^{a(x)},\phi^{{\bar A}(x)})$.
The higher dimensional metric is a functional of $\phi^{A(x)}$.
Performing the Kaluza-Klein split,
\be
G_{\,\,\,\,\,\,\,AB}^{\mu \nu }  = \left( \begin{array}{l}
 {\gamma^{\mu \nu }}_{ab}  + {A_a}^{\bar A} {A_b}^{\bar B} 
 {{\bar G} ^{\mu \nu }}_{\bar A\bar B} \,,\,\,\,\,\,\,\,\,\,\,\,
 {A_a}^{\bar B} {{\bar G}^{\mu \nu }}_{\bar A\bar B} \, \\ 
 \,\,\,\,{A_b}^{\bar B} {{\bar G}^{\mu \nu }}_{~\bar A\bar B} 
 \,,\,\,\,\,\,\,\,\,\,\,\,\,\,\,\,\,\,\,\,\,\,\,\,\,\,\,
 \,\,\,\,\,\,\,{{\bar G}^{\mu \nu }}_{~\bar A\bar B} \,\,\, \\ 
 \end{array} \right) ,
\lbl{q5.35}
\ee
where for simplicity we have omitted the index $(x)$, we find,
\be
I_\phi   = \frac{1}{2}\partial _\mu  \phi ^{a(x)} \partial _\nu  \phi ^{b(x')}
 \gamma _{\,\,\,\,\,\,\,\,a(x)b(x')}^{\mu \nu }  
 + \frac{1}{2}\partial _\mu  \phi _{{\bar A}(x)} 
 \,\partial _\nu  \phi _{{\bar B}(x)}
  \,{\bar G}^{\mu \nu \bar A (x) \bar B (x)} .
\lbl{q5.36}
\ee
Identifying $\phi^{a(x)} \equiv \varphi^{a(x)}$, and
denoting $\frac{1}{2}\partial _\mu  \phi _{\bar A (x)} 
\,\partial _\nu  \phi _{\bar B (x)}
  \,{\bar G}^{\mu \nu \bar A (x) \bar B (x)} = - U[\varphi]$,
we obtain the action (\ref{q5.33}).

Thus, the field action (\ref{8.1}) is embedded in a higher dimensional action
with a metric $G_{\,\,\,\,\,\,A(x) B(x)}^{\mu \nu }$ in field space. A question
arises as to which field space metric to choose. The lesson from general relativity
tells us that the metric itself is dynamical. Let us therefore assume that this
is so in the case of field theory\,\ci{PavsicBook} as well. Then (\ref{q5.34})
must be completed by a kinetic term, $I_G$, for the field space
metric\,\ci{PavsicBook}. The total action is then
\be
  I[\phi ,G] = I_\phi   + I_G .
\lbl{q5.37}
\ee
According to such dynamical principle, not only the field $\phi$, but
also the metric $G_{\,\,\,\,\,\,A(x) B(x)}^{\mu \nu }$ changes with the
evolution of the system. This implies that also the potential $U[\varphi]$
of eq.\,(\ref{q5.33}) changes with evolution, and so does the potential
$V(\varphi)$, occurring in eq.\,(\ref{8.1}).

Let us assume that the action (\ref{q5.37})   describes the whole
universe\footnote{Of course, such a model universe is not realistic, because our
universe contains fermions and accompanying gauge fields as well.}.
Then $|\psi (t) \rangle$ of eq.\,(\ref{8.7}) contains everything in such
universe, including observers. There is no external observer,
${\cal O}_{\rm ext}$,
according to whom the coefficients $\langle p_1 ...p_n|\psi (t) \rangle$
in eq.\,(\ref{8.7}) could be related to the probability densities
$|\langle p_1 ...p_n|\psi (t) \rangle|^2$ of finding the system in an
$n$-particle states with momenta $p_1$, ..., $p_n$, $n=0,1,2,3,...,\infty$.
There are only inside observers, ${\cal O}$, incorporated within 
appropriate multiparticles states $|p_1 ...p_n \rangle$, $n=0,1,2,3,...,\infty$,
of the ``universal" state $|\psi (t) \rangle$. According to the Everett
interpretation of quantum mechanics, all states,  $|0 \rangle$,
$|p_1 p_2 \rangle$,..., $|p_1 ... p_n \rangle$, ..., $n=0,1,2,...,\infty$,
in the superposition (\ref{8.7}) actually occur, each in a different world.
The Everett interpretation is now getting increasing support among
cosmologists (see, e.g., Ref.\,\ci{Tegmark,Zeh}).
For such an
inside observer, ${\cal O}$, there is no instantaneous vacuum decay into
infinitely many particles. For ${\cal O}$, at a given time $t$, there exists a
configuration $|P \rangle$ of $n$-particles (that includes ${\cal O}$
himself), and a certain field potential
$V(\varphi)$ (coming from ${G^{\mu \nu}}_{A(x)B(x)}$).
At some later time, $t+\Delta t$, there exists a slightly different
configuration $|P' \rangle$ and potential $V'(\varphi)$, etc.
Because ${\cal O}$ nearly continuously measures the state $|\psi (t) \rangle$ of
his universe, the evolution of the system is being ``altered", due to the notorious
``watchdog effect" or ``Zeno effect"\ci{Sudarshan} of quantum mechanics (besides being altered by
the evolution of the potential $V(\varphi)$). The peculiar behavior of a quantum
system between two measurements has also been investigated
in Refs.\,\ci{Aharonov,Aharonov1}.
\end{quotation}

In Discusion of the same reference \ci{PavsicFirenze} it is then written:

\begin{quotation}

After having investigated how the theory works on the examples of the
classical and quantum pseudo Euclidean oscillator, we considered quantum
field theories. If the metric of field space is neutral, then there occur
positive and negative energy states. An interaction causes transitions
between those states. A vacuum therefore decays into a superposition of
states with positive and negative energies. Because of the vast phase space
of infinitely many particles, such a vacuum decay is instantaneous---for
an external observer. But we, as a part of our universe, are not external
observers. We are internal observers entangled with the ``wave function''
of the rest of our universe. According to the Everett interpretation of quantum
mechanics, we find ourselves in one of the branches of the universal
wave function. In the scenario with decaying vacuum, our branch can consist
of a finite number of particles. Once being in such a branch, it is improbable
that at the next moment we will find ourselves in a branch with infinitely
many particles. For us, because of the ``watchdog effect" of quantum mechanics,
the evolution of the universal wave function relative to us is frozen to
the extent that instantaneous vacuum decay is not possible. 

\end{quotation}

Another important point is that the field potential in the action (\ref{8.1})
need not be fixed during the evolution of the universe, but can change.
Also, a realistic potential is not unbounded, it should be bounded from
below and from above, which then presumably gives stability as we have
shown that it is the case in the classical theory (see Sec.\,4.2).
Moreover, as pointed by Carroll et al.\,\ci{Carroll}, decay rates of the processes
with negative energy particles (``phantoms" in their terminology) would not diverge
if there were a cutoff to the phase space integrals. Maximum momenta 
imply minimum length, and in Ref.\,\ci{PavsicFirenze} it was pointed out
that such a finite system (though with many degrees of freedom) oscillates
between the ``vacuum" state and the very many (but not infinitely many)
particle state.

A realistic description of our universe must include fermions as well. A well known
fact is that fermions are just particular Clifford numbers (algebraic spinors or 
Clifford aggregates)\,\ci{PavsicSpinorInverse}--\ci{BudinichFock2}. An element
of the Clifford algebra generated by the basis
vectors of 4-dimensional spacetime can be represented by a $4 \time 4$ matrix,
and spinors are just elements of one column (an ``ideal" of a Clifford
algebra). Because there are four columns, there
are four kinds of spinors. In Ref.\,\ci{PavsicFirenze} we find the following discussion
(after a lengthy formalism supporting it):

\begin{quotation}

Description of our universe requires fermions and accompanying gauge fields,
including gravitation. According to the Clifford algebra generalized
Dirac equation---Dirac-K\"ahler equation\,\ci{DiracKahler}--\ci{Spehled}---there are four sorts of the
4-component spinors, with energy signs a shown
in eq.\,(111) [of Ref.\,\ci{PavsicFirenze}]. The vacuum of
such field has vanishing energy and evolves into a superposition of positive
and negative energy fermions, so that the total energy is conserved. 
A possible scenario is that the
branch of the superposition in which we find ourselves, has the sea of
negative energy states of the first and the second, and the sea positive
energy states of the third and forth minimal left ideal of $Cl(1,3)$.
According to ref.\,\ci{PavsicInverse,PavsicMoscow}, the former states are
associated with the
familiar, weakly interacting particles, whereas the latter states are
associated with mirror particles, coupled to mirror gauge fields, and thus
invisible to us. According to the field theory based on the Dirac-K\"ahler
equation, the unstable vacuum could be an explanation for Big Bang.

\end{quotation}

Although the concept of the Dirac vacuum as a sea of negative energy
particles is nowadays considered as obsolete, it finds its natural
place within the Clifford algebra description of spinors and spinor
fields\,\ci{PavsicSpinorInverse}--\ci{PavsicBookLicata}.
If a higher derivative field theory describes our
universe, then as a possible scenario it predicts that an initial ``bare" vacuum state
evolved into the present state, which involves the sea of negative energy fermions
of the 1st and 2nd ideal (column) of Cl(1,3), and the sea of positive energy fermions
of the 3rd and 4th ideal, so that their total energies sum to zero. The initially
``bare" vacuum thus decayed into infinitely many particles not completely,
but only partially, because half of the available fermion states remained finite,
namely the positive energy states of the 1st and 2nd (that form our
visible universe, including us), and the negative
energy states of the 3rd and 4th ideal (that according to
Ref.\,\ci{PavsicSpinorInverse,PavsicMoscow,PavsicBookLicata})
form mirror particles. Our sector of particles has remained finite, because we live
in it and we observe (or measure) it. In other words, we are composed of those particles
and we observe the same kind of particles in our environment and measure (at least
implicitly) their momenta. But in this scenario we have not measured the momenta
of those other kinds of fermions which then formed the sea. So people when talking
about an (instantaneous) vacuum decay were only ``half right". Vacuum indeed
(presumably instantaneously) decayed into infinitely many particles, but only into
half of the available four kinds of (fermionic) particles.

\section{Conclusion}

We have reviewed various approaches in the literature to the problem of
negative energies that occur in higher derivative theories, a toy model for them
being the Pais-Uhlenbeck oscillator. We have pointed out to the well-known fact
that negative energies themselves are not problematic, if there are no interactions
between positive and negative energy degrees of freedom. The problems are
expected to arise in the presence of interactions. Therefore, the approaches in which
the authors show how the description of the PU oscillator in terms of an indefinite
Hamiltonian can be replaced with a positive definite Hamiltonian, do not solve
the problem. Namely, such a reformulation of a free PU oscillator does not work
for an interacting PU oscillator. For instance, if a self-interacting PU oscillator is
attempted to be reformulated in terms of a positive definite Hamiltonian, the
Lagrangian acquires additional non linear terms that are not present in the
original Lagrangian. Therefore, an interacting PU oscillator as a higher derivative
system has to be described in terms of the Ostrogradski or an equivalent formalism
that contains negative energies. It was shown that if the  potential is unbounded
either from below or from above, then such systems are not necessarily unstable;
they can be either stable or unstable, depending on the values of the coupling constant 
and the initial velocity. However, when quantized, the system becomes unstable because
of the tunneling effect, even if it was stable in the classical case.

But unbounded potentials are not physically realistic. For instance, a harmonic oscillator
is an idealization, not realized in nature. An actual oscillator always has a potential that is bounded.
In the case of systems whose energies are positive, the potential is required to be
bounded from bellow, but usually it is taken for granted that it need not be bounded from
above, despite that a physically realistic system is bounded form above as well. When
considering higher derivative theories that contain negative energies, if we assume the presence
of an unbounded potential, then the systems described by such theories are not stable. As we
mentioned, they can be stable within certain ranges of initial conditions and coupling constant,
but they cannot be absolutely stable, i.e., stable for all values of those parameters.
In this review we pointed out that a physically realistic potential has to be bounded not only from
bellow, but also from above. Then even the systems whose energies can be positive or negative,
are absolutely stable. When such systems are quantized, they remain stable.
An alternative approach to the interacting higher derivative systems, including the PU oscillator
has, been considered in Refs. \cite{Kaparulin:2014vpa}--\ci{Kaparulin:2015uxa} 

The issue regarding the infinite decay rate of vacuum due to the infinite phase space in higher derivative
fields theories has also been clarified. The integration over phase space has to be performed
when momenta of decaying products are not measured. If they are measured, then the integration
does not run over the entire infinite phase space, but only over a small finite part of it, and therefore
the decay rates are not infinite. Assume that our universe is governed by a higher derivative
field theory. Then we observers living in such a universe are part of one of the infinitely
many components of such a decayed vacuum, and we in fact  implicitly measured momenta
of the particles in our universe, therefore for us there was no instantaneous vacuum decay. Moreover,
if there is a momentum cutoff due to the minimal length, the phase space is not infinite anyway.
We conclude that higher derivative theories are physically viable with many important
physical implications for further development of quantum gravity and its unification with the
rest of physics.

\end{document}